\documentclass[iop,revtex4-1,numberedappendix]{emulateapj}%iop,twocolumn...
\usepackage{subfigure,graphicx}
\usepackage{url}
\usepackage[bookmarks=false]{hyperref}
\usepackage{longtable}
\usepackage{appendix}
\usepackage[T1]{fontenc}
\usepackage{graphicx}
\usepackage[outdir=./]{epstopdf}
\usepackage{graphics}
\usepackage{placeins}
\usepackage{float}
\usepackage{xspace}
\usepackage{afterpage}
\hypersetup{
  colorlinks= true, %Colors links instead of ugly boxes
  urlcolor  = black, %Color for external hyperlinks
  linkcolor = red, %Color of internal links
  citecolor = blue %Color of citations
}

\newcommand{\herschel}{\textit {Herschel}\xspace}

\newcommand{\iras}{\textit {IRAS}\xspace}
\newcommand{\iso}{\textit {ISO}\xspace}
\newcommand{\ci}{[\ion{C}{1}]\xspace}
\newcommand{\cii}{[\ion{C}{2}]\xspace}

\newcommand{\oi}{[\ion{O}{1}]\xspace}

\newcommand{\nii}{[\ion{N}{2}]\xspace}
\newcommand{\hi}{\ion{H}{1}\xspace}
\newcommand{\hii}{\ion{H}{2}\xspace}

\shorttitle{Herschel Spectroscopy of Early Type Galaxies}   % Short form of title for running header in final publication
\shortauthors{Lapham, Young, and Crocker} % Short author list for running header in final publication

\begin{document}

\title{Herschel Spectroscopy of Early Type Galaxies}

\author{Ryen Carl Lapham and Lisa M. Young}
\affil{Physics Department, New Mexico Institute of Mining and Technology, 801 Leroy Place, Socorro, NM 87801; ryen.lapham@student.nmt.edu, lyoung@physics.nmt.edu}

\and

\author{Alison Crocker}
\affil{Physics Department, Reed College, Portland, OR 97202; crockera@reed.edu}

%\author{Lisa Young}
%\affil{New Mexico Institute of Mining and Technology}
%\affil{801 Leroy Place, Socorro, NM 87801, USA}
%\email{lyoung@physics.nmt.edu}

\begin{abstract}
We present \herschel spectroscopy of atomic lines arising in photodissociation regions as well as ionization regions of nearby early-type galaxies (ETGs), focusing on the volume-limited Atlas3D sample. Our data include the \cii, \oi, and \nii 122 and 205 \micron\ lines, along with ancillary data including CO and \hi maps. We find ETGs have \cii/FIR ratios slightly lower than spiral galaxies in the KINGFISH sample, and several ETGs have unusually large \nii 122/\cii ratios. The \nii 122/\cii ratio is correlated with UV colors and there is a strong anti-correlation of \cii/FIR with NUV-K seen in both spirals and ETGs, likely due to a softer radiation field with fewer photons available to ionize carbon and heat the gas.  The correlation thus makes a \cii deficit in galaxies with redder stellar populations.  The high \nii 122/\cii (and low \cii/FIR) line ratios could also be affected by the removal of much of the diffuse, low density gas, which is consistent with the low \hi/H$_2$ ratios.  \cii is now being used as a star formation indicator, and we find it is just as good for ETGs as in spirals. The \cii/CO ratios found are also similar to those found in spiral galaxies. Through use of the \nii 205 \micron\ line, estimates of the percentage of \cii emission arising from ionized gas indicate a significant portion could arise in ionized regions.

\end{abstract}

\keywords{galaxies: elliptical and lenticular, cD --- galaxies: ISM } %Single piece of text. Get list of appropriate keywords from publisher's website.

%%%%---------------- Introduction --------------------------
\section{INTRODUCTION}

The atomic and ionic fine-structure lines \cii (158 $\mu$m) and \oi (63 $\mu$m) are the dominant cooling lines of the neutral interstellar gas, and therefore they are important for understanding the heating and cooling balance in this phase of the interstellar medium. The \nii (122, 205 $\mu$m) lines are also prominent FIR tracers of ionized gas.  These lines provide diagnostics for inferring physical conditions of the gas such as temperatures, densities, and radiation field strengths by comparing with models of photodissociation regions \citep{KWH06, PW08}. They are also used as tracers of the star formation rate, similar in spirit to the optical nebular lines, but with lower extinction \citep{Zhao2016}. And as the FIR lines are convenient for high resolution millimeter interferometry at redshifts $\gtrsim$ 3, they are fast becoming workhorses for the study of high redshift galaxy evolution.  In the context of understanding the high redshift universe, it should be helpful to understand the behavior of these lines in as many different types of local galaxies as possible. There is some previous work on the lines using Infrared Space Observatory (\iso) data on a wide variety of galaxy types (e.g. \citet{Mal01} and \citet{Brauher}), but the launch of the \herschel \footnote{Herschel is an ESA space observatory with science instruments provided by European-led Principal Investigator consortia and with important participation from NASA.} satellite provided the opportunity to observe these lines in more sources and at higher resolution than was previously possible.  

Only a handful of early-type galaxies have previous \cii and \oi observations, and hence little is known about the neutral interstellar medium (ISM) in these galaxies. \citet{Mal00} observed four early type galaxies (two elliptical and two lenticular) with the \iso, and detected all four galaxies in \cii, but only two in \oi.  They found that \cii/FIR was lower than typical values found in other samples by a factor of 2-5, and postulated that a softer UV radiation field was the explanation for this, corroborated by a low \cii/CO ratio in NGC 5866.

\citet{Mal01} carried out a large study with \iso\ on 60 normal, star-forming galaxies on a variety of fine structure lines including \cii 158 \micron, \oi 63 and 145 \micron, \nii 122 \micron, [\ion{O}{3}] 52 and 88 \micron, and [\ion{N}{3}] 57 \micron. \cii/FIR was found to decrease with dust temperature due to the dust grains becoming positively charged, decreasing the efficiency of photoelectric ejection.  This decrease was stronger than that found for ETGs with softer radiation fields.  The \oi/\cii ratio was lower than PDR models predicted, implying a contribution of \cii emission from ionized gas, but \oi/\cii correlated well with 60 \micron/100 \micron\ (a proxy for dust temperature) as predicted.  PDR models found the average UV radiation strength to be $10^2<G_0<10^{4.5}$ and the gas density to be $10^2<n<10^{4.5}$ cm$^{-3}$, with $G_0$ $\alpha$ $n^{1.4}$.

\citet{Brauher} carried out a larger study of FIR emission lines on 227 galaxies (46 resolved, 181 unresolved) observed by the \iso.  The trend of \cii/FIR decreasing with dust temperature was confirmed for all morphological types.  ETGs in the sample have values similar to the other galaxies.  Unlike \cii/FIR, which decreases with dust temperature, \oi/FIR is found to have no trend with 60 \micron/100 \micron\ for all galaxies, so as the dust temperature increases, \oi takes on a larger role in cooling the ISM. They find the \nii122/FIR ratio to decrease in a similar way to the \cii/FIR ratio for spirals and irregulars, but do not have enough detections to make a conclusion on other morphologies. The  \oi/\cii line ratio increases with dust temperature for all galaxy types, and \oi becomes the primary ISM coolant around 60 \micron/100 \micron\ $\geq$ 0.8.  The \nii122/\cii ratio is roughly constant for all types of galaxies, and they report a median value of 0.11 for galaxies with detections of both lines.

This paper provides a FIR line-based characterization of the warm and cool ISM of the largest sample of ETGs to date, allowing comparison with both normal spiral galaxies and the central galaxies of cooling flow clusters. In Section \ref{sec:data} the selection criteria are explained, along with the choices for the observation settings, the data reduction process, and the methods for measuring line fluxes.  The data analysis is explained in Section \ref{sec:analysis}, including a discussion of line ratios, separating the sources of \cii emission, and different uses for the \cii line.  Conclusions are presented in Section \ref{sec:conclusion}.

%%%%---------------- Data --------------------------
\section{DATA}
\label{sec:data}

\subsection{Sample Selection}
\label{sec:selection}

The galaxies observed here were selected primarily as the most CO-bright and FIR-bright members of the Atlas3D sample \citep{Cappellari}.  The Atlas3D sample was developed as a complete volume-limited set of early-type galaxies brighter than $M_{K}$ = -21.5 (stellar masses greater than about $10^{9.9}$ M$_{\mathrm{\sun}}$); they have distances out to 42 Mpc,  declinations between -4$^{\circ}$ and +64$^{\circ}$, and galactic latitudes > 15$^{\circ}$.  In this context the term ``early-type'' refers to galaxies lacking spiral structure, and with masses large enough to avoid irregulars and dwarf spheroidal galaxies.  No color cut was made on the sample, so it includes a few relatively blue galaxies.  The core of the Atlas3D project involved integral-field optical spectroscopy of the central regions of the 260 galaxies that meet all these criteria, and observations of the cold atomic and molecular gas were also carried out (\citet{Serra}; \citet{Young2011}).  

\begin{figure}
\epsscale{1.2} %Override for scale, in decimal units, e.g., 0.80
\begin{tiny}
\label{•} 
\end{tiny}\plotone{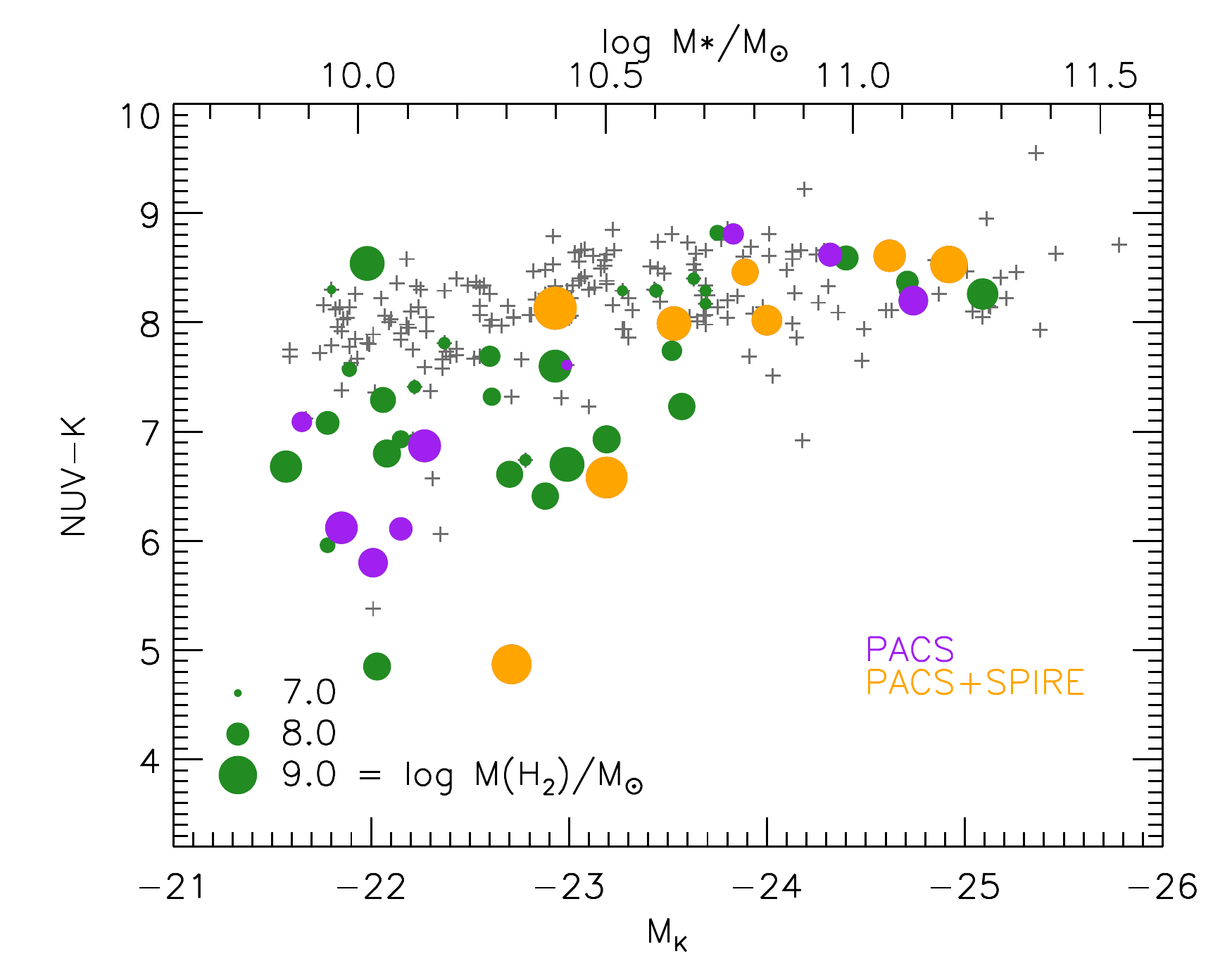} %Plots one file, image width scaled to text width
\caption{Properties of the sample galaxies relative to the full Atlas3D sample. Circle sizes indicate molecular gas content. Purple circles represent galaxies observed with PACS; yellow circles represent galaxies observed with both PACS and SPIRE. Green circles represent Atlas3D galaxies with molecular gas, but not selected for Herschel observations, while crosses indicate Atlas3D galaxies without detected CO.  The red sequence of ETGs is prominent at NUV-K $\sim$8-9.}
\label{CMD}
\end{figure}

Of the 23 Atlas3D galaxies brightest in FIR flux, we observed 18 with PACS and the KINGFISH project observed 2 \citep{Kenn2011}.  Of the 14 brightest in $^{12}$CO J=1-0 (as well as FIR), we selected 8 for SPIRE spectroscopy and obtained another one from the archive. This was a conservative approach to make detections more likely, and our results show that fainter galaxies would likely have been detected had they been chosen. An optical color-magnitude diagram of the sample galaxies is shown in Figure \ref{CMD}.  This set includes field galaxies and members of the Virgo Cluster; some are \hi-rich, and some (even field galaxies) are extremely \hi-poor, with the \hi/H$_2$ ratio ranging from $\sim$ 0.001 to 10; some are found on the red sequence (NUV-K $\sim$ 8.5), whereas others are as blue as typical spirals (NUV-K $\sim$ 5); some have dynamically relaxed cold gas disks whereas others show clear signs of recent accretion or disturbance.  Some have AGN activity.  Thus, the sample probes a wide variety of parameters that might affect the properties of their ISM.

\subsection{Observation and Reduction}
\label{sec:observations}

\subsubsection{PACS Observations}
\label{sec: PACS observations}

Eighteen galaxies were observed with the PACS spectrometer \citep{Pog} and a subset of 8 had additional observations with SPIRE \citep{Griff} under the project OT1\_lyoung\_1. Both instruments were part of the \herschel space-borne observatory \citep{Pil10}. Our data obtained with the PACS spectrometer include the \cii 158 $\mu$m, \oi 63 $\mu$m, and \nii 122 $\mu$m fine-structure lines.  Optical images of each galaxy with an outline of the PACS field of view and \cii contours can be seen in Figure \ref{optical}. We also make use of PACS observations for NGC 1266 and NGC 5866 from the KINGFISH project KPOT\_rkennicu\_1.  All lines were observed in the PACS LineSpec mode except for the \oi lines in NGC 3607, NGC 3626, NGC 3665, NGC 4429, NGC 4435, NGC 4459, and NGC 4526 which were observed in the RangeSpec mode.   RangeSpec mode was used when there were concerns that the wavelength range available in the standard LineSpec mode at 63 $\mu$m would not be large enough for their broad lines.  This is based on previously known CO linewidths \citep{Young2011}.   The \oi line was not observed for UGC01676, and the wrong velocity was used for NGC 3665 leading to a non-centered detection of the \oi line.  The observations were done in LineSpec ``faint line'' mode or (as appropriate) RangeSpec with Nyquist sampling. The spectral resolutions for the \cii, \nii, and \oi lines are 237.8 km/s, 291.2 km/s, and 87.5 km/s, respectively.

Our observations were done as small raster maps in Chop-Nod mode with a large throw (6 \arcmin). \cii and \nii used 2x2 rasters with a step of 4.5\arcsec.  The \oi LineSpec observations were done as 3x3 rasters with a step of 3\arcsec, and the \oi RangeSpec ones were done as 2x2 rasters with a step of 4.5\arcsec.  Observations were done as raster mapping to fully sample the point spread function and provide information on the extent and kinematics of the gas.  PACS spectroscopy has an angular field of view of 47\arcsec\ x 47\arcsec\ and pixel sizes of 9\arcsec\ x 9\arcsec.  

Our PACS data were reduced using the \herschel Interactive Processing Environment (HIPE) version 14 \citep{Ott10}.  The built-in ChopNod pipeline was used to calibrate the flux and wavelengths, as well as mask bad pixels (outliers, glitches, saturated or noisy pixels). To avoid correlated noise, our pipeline wavelength gridding uses values of 2 for oversample and 1 for the upsample parameters.  The resulting data cubes were then further processed in HIPE's spectrum explorer.  The continuum was subracted using a degree two polynomial model based on manually selected regions, avoiding the noise at the edges of the spectra.

\subsubsection{SPIRE Observations}
\label{sec: SPIRE observations}

In addition to the PACS observations, a subset of nine galaxies were observed with the SPIRE instrument. Based on CO images from \citet{Alatalo}, the molecular emission was expected to be mostly or entirely within the beam of the central SPIRE detector, so the observations are single pointings in the high spectral resolution mode.  Many CO transitions are seen in the SPIRE spectra, along with two \ci  transitions, a few H$_{2}$O and HCN transitions, and occasional HCO lines.  The majority of the CO and \ci lines were detected in almost all of the galaxies. In this work, only the \nii 205 line is used from the SPIRE spectra. The 205 \micron\ data from SPIRE is only used for our ETGs.  For the KINGFISH galaxies, the 205 \micron\ line fluxes instead come from PACS data.  In Appendix \ref{sec:kfcompare} we discuss the good agreement between the PACS and SPIRE \nii 205 \micron\ line fluxes using 6 KINGFISH galaxies with measurements from both instruments.  All other lines observed with SPIRE will be discussed in a subsequent paper. The details of the observations for each galaxy can be found in the Herschel Science Archive (HSA).  SPIRE photometry observations are available for all of the SPIRE spectroscopic targets to assist with calibration.

The SPIRE data were reduced using HIPE 14's built in Single Pointing user pipeline. After the standard pipeline reduction the spectra were processed with the Semi-Extended Flux Correction Tool (SECT).  The SECT parameters are discussed below.

\subsection{Comparison of Cooling Line Emission Distribution}
\label{sec:distribution}

We begin by estimating how much of the FIR line emission may be lost outside of the PACS field of view by determining how much of the 160 $\mu$m continuum is emitted outside this field of view.  PACS 160 $\mu$m continuum images were overlaid with the PACS spectral footprint, and the flux within the footprint was compared to the total flux contained within the galaxy. PACS photometry was available in the HSA for NGC 1266, 3626, 3665, 4459, 4526, 4710, 5866, and 7465.  The worst cases were NGC 4710, which had 88.7\% of the 160 $\mu$m emission in the PACS field of view, and NGC 4526, which contained 95.8\% of the continuum emission in the PACS field of view. The continuum emission for the remaining galaxies was completely inside the PACS spectral footprint.  If \cii is distributed like the 160 $\mu$m continuum, then we expect the PACS field of view to contain $\geq$ 90\% of the total \cii flux.

%removed sentence%NGC 4526 required a 60\arcsec\ x 75\arcsec\ region to capture all of the continuum flux, while NGC 4710 required a 90\arcsec\ x 60\arcsec\ region.  

Additionally, position-velocity slices for \cii, \oi, and \nii 122 micron were utilized along with those of CO(1-0) to examine whether the atomic gas is spatially co-located with the molecular gas, as well as sharing kinematic properties. In most cases the atomic gas follows the molecular gas, as in Figures \ref{pvslice1}, \ref{pvslice2}, and \ref{pvslice3}.  This is reassuring for galaxies with faint \oi emission such as NGC 3489, 4429, or 3607.  NGC 3626 may be a case where \oi is more compact than CO.  In some cases \cii is more spatially extended than CO.  This is conspicuous in NGC 1222, seen in Figure \ref{pv1222}, where the \cii likely follows the sum of atomic and molecular gas, but could be the case in NGC 4694, 7465, and IC 1024 as well. The line widths of the PACS lines are also similar to those found from CO, though the resolution for the CO is much better and can more accurately describe the motion of the gas.

\subsection{Measuring Line Fluxes}
\label{sec:lineflux}

In Table \ref{fluxtable} we present the fluxes and uncertainties obtained by summing the spectral data from the entire PACS image as well as the fluxes obtained in an aperture of radius 15\arcsec\ centered on the galaxy nucleus (the brightest pixel).  In both cases the data cubes were spatially integrated to produce a single spectrum from which a total flux could be extracted using a Gaussian model.  The pixels are summed directly without the use of weighting.  The spectra produced with 15\arcsec\ apertures can be seen in Figure \ref{spectra}. For the 15\arcsec\ apertures, Enclosed Energy Fractions (EEFs) were used to correct the spectra for missing emission scattered out of the aperture by the point spread function; these corrections assume the target is a point source, and while our galaxies are not true point sources, they are centrally peaked.  Uncertainties in the line fluxes were calculated using the residuals from the Gaussian model.  A single-channel standard deviation $\sigma_{1}$ was calculated and summed across the channels that span the full width of the line, giving a total uncertainty in the sum of $\sqrt{N} \sigma_{1} \delta v $ , where $N$ is the number of channels spanning the width of the line and $\delta v$ is the channel width. It should be noted there is an additional $\pm$ 30\% absolute calibration uncertainty for PACS spectroscopy as detailed in Section 4.10 of the \href{http://herschel.esac.esa.int/Docs/PACS/html/pacs_om.html}{PACS Observer's Manual}. 

Fluxes from the two apertures are presented because they are useful in different situations.  In calculating a line/continuum ratio such as [CII]/FIR, our FIR fluxes come from low resolution IRAS, Spitzer, or AKARI data so the largest available line aperture is desired. However, for the fainter lines, the outer parts of the PACS field of view are very noisy and the nominal fluxes measured in the full image can be less than fluxes measured in a smaller aperture.  In addition, for internal PACS line ratios it is desirable to match the apertures used in each line.  For these ratios we use the matched 15\arcsec\ apertures.  The choice of the 15\arcsec\ radius is based on inspection of the position-velocity diagrams (Figures \ref{pvslice1}, \ref{pvslice2}, and \ref{pvslice3}) and a curve-of-growth analysis; it is large enough to capture most of the flux in the PACS fields, to provide robust detections and to have minimal EEF corrections ($\leq$ 10\%), but small enough to avoid most of the noisy regions around the field edges.  For our targets the 15\arcsec\ radius corresponds to $\sim$ 1 to 2 kpc.

Galaxies NGC 1222, 1266, 4429, and 5866 have previous observations made by the Long Wavelength Spectrometer aboard \iso\ \citep{Brauher}. For NGC 1222, our \cii flux is 24\% greater, our \oi flux is 86\% of their value, and the \nii values agree within 10\%.  For NGC 1266, our \cii flux is 14\% greater and the \nii flux is consistent with their upper limit. For NGC 4429 only a \cii observation was available, but it is within 5\% of our value.  Finally, for NGC 5866 our PACS line fluxes obtained within a 15\arcsec\ aperture are consistent with the \iso\ data. The \cii flux is 13\% greater than what was previously found, and the \nii flux is 48\% less than the previous value. However, both \nii detections are weak ($\sim$ 3.5 $\sigma$), so they are consistent given the beam sizes and low S/N.

Before measuring the fluxes in our SPIRE spectra, we apply a correction factor determined by the semi-extended correction tool (SECT) in HIPE. The SECT attempts to extrapolate to a total source flux, correcting for the fact that the central detector beam may not fully cover the source and that the beam size is strongly wavelength dependent.  When using the SECT on SPIRE spectra, we prefer to specify the source size, position angle, and eccentricity, rather than allowing the SECT to optimize the source size based on the overlap region between SLW and SSW spectra.  The reason for this is that the two parts of the spectrum have random continuum offsets (with respect to each other) on the order of 0.3-0.4 Jy \citep{Hopwood}, and for these weak sources that is a large proportion of the flux in the overlap region.  The spatial and kinematic CO comparisons--particularly \nii 122 vs CO shown in Figure \ref{pvslice2}--suggest that the CO sizes offer good estimates to use in the SPIRE SECT when measuring the \nii 205 fluxes. We also prefer not to force our spectrum to match the photometry, but instead use external knowledge about the probable emission size and use the photometry as a sanity check.  The source variables were chosen using two dimensional Gaussian models of the CO(1-0) emission. For NGC 2764 and 5866 the corrections based on CO sizes clearly underestimate the photometry, so larger source sizes based on two-dimensional Gaussian fits to the SPIRE 250\micron\ images were also used. We make use of both \nii 205 line fluxes for these two galaxies in a ``worst case'' illustration of the magnitudes of the extrapolation uncertainties.  A Gaussian reference beam with FWHM = 44.15\arcsec\ was used to obtain a flux matching the 47\arcsec\ x 47\arcsec\ field of view for the PACS observations.

After being reduced in the standard pipeline and processed with the SECT, the SPIRE spectra were fit using HIPE's built in Spectrometer Line Fitting script. The spectra before and after the SECT are seen in Figure \ref{SPIREspectra}.  The unapodized spectra were used in the fitting routines, and two components were used to model the emission. Sinc line profiles were used for all galaxies except NGC 4526, which was fit using a sinc-gauss line profile.  Uncertainties were calculated using the residuals from the model in the same manner as the PACS spectra. Fluxes and uncertainties for the SPIRE data, as well as the correction factor given by the SECT at 205 $\mu$m can be found in Table \ref{SPIREdata}.  \href{http://herschel.esac.esa.int/Docs/SPIRE/html/spire_om.html}{The SPIRE Handbook version 2.5} recommends summing the absolute calibration uncertainty (4\%) and relative calibration uncertainty (1.5\%) directly for a conservative total estimate of the calibration uncertainty of 5.5\%.   The calibration uncertainties have not been applied to the line flux uncertainties listed in the tables for PACS or SPIRE.

Systematic uncertainties due to the source size extrapolation corrections for our galaxies are expected to be similar to or larger than the absolute calibration uncertainties. Comparison to \citet{Kam2016} suggests these systematic uncertainties are on the order of $\sim$20-30\%. To obtain flux measurements they first use the 250 \micron\ continuum for a size estimate, fixing the SLW/SSW overlap, but this often makes the spectra higher than the photometric data, so they multiply down by a polynomial to ensure the spectrum matches the photometry.  We use a different approach, but  our SPIRE \nii 205 line fluxes are in very good agreement with their values.  NGC 2764, NGC 3665, NGC 4526, and NGC 5866 are almost exact matches, within a few percent, while the remaining galaxies agree within 20-30\%.  The NGC 2764 and 5866 fluxes found using the 250 \micron\ sizes are 74\% and 72\% greater than the \citet{Kam2016} values.

\setcounter{table}{0}
\begin{deluxetable*}{ llcllllllllll }
\tablecaption{Line Fluxes and Uncertainties}
\tablehead{
 \colhead{} & \colhead{} & \colhead{} & \multicolumn{2}{c}{\cii 158} & \multicolumn{2}{c}{\oi 63} & \multicolumn{2}{c}{\nii 122} & \multicolumn{2}{c}{FIR} \\ \colhead{Galaxy} & \colhead{Int. Area} & \colhead{Size (kpc)} & \colhead{Flux} & \colhead{$\sigma$} & \colhead{Flux} & \colhead{$\sigma$} & \colhead{Flux} & \colhead{$\sigma$} & \colhead{Flux} & \colhead{$\sigma$}  } 
\startdata

IC 0676 & 15 \arcsec & 1.8 & 6.674 & 0.062 & 4.131 & 0.062 & 1.089 & 0.058 & 1692 & 17\\
       & full image & & 7.938 & 0.072 & 4.04 & 0.27 & 1.145 & 0.076 & 1692 & 17\\
IC 1024 & 15 \arcsec & 1.8 & 15.010 & 0.038 & 5.632 & 0.056 & 0.986 & 0.039 & 2316 & 17\\
       & full image & & 18.715 & 0.077 & 7.10 & 0.35 & 0.910 & 0.082 & 2316 & 17 \\
NGC 1222 & 15 \arcsec & 2.4 & 30.02 & 0.19 & 23.379 & 0.072 & 1.191 & 0.036 & 6187 & 19\\
        & full image & & 34.73 & 0.13 & 25.19 & 0.25 & 1.202 & 0.070 & 6187 & 19\\
NGC 1266 & 15 \arcsec & 2.2 & 6.92 & 0.18 & 4.94 & 0.30 & 1.15 & 0.17 & 8481 & 283\\
        & full image & & 7.24 & 0.33 & 5.15 & 0.89 & \nodata & \nodata & 8481 & 283\\
NGC 2764 & 15 \arcsec & 2.9 & 14.693 & 0.066 & 5.64 & 0.17 & 1.186 & 0.036 & 2103 & 85\\
       & full image & & 19.225 & 0.053 & 7.40 & 0.72 & 1.457 & 0.071 & 2103 & 85\\
NGC 3032 & 15 \arcsec & 1.6 & 6.995 & 0.024 & 2.271 & 0.070 & 0.946 & 0.022 & 1976 & 93\\
        & full image & & 7.738 & 0.049 & 2.29 & 0.20 & 1.023 & 0.080 & 1976 & 93\\
NGC 3489 & 15 \arcsec & 0.9 & 1.643 & 0.031 & 0.459 & 0.075 & 0.181 & 0.020 & 1468 & 43\\
        & full image & & 2.65 & 0.12 & $<$ 0.903 & \nodata & \nodata & \nodata  & 1468 & 43\\
NGC 3607 & 15 \arcsec & 1.6 & 1.551 & 0.055 & $<$ 0.494 & \nodata & 0.622 & 0.069 & 1305 & 58\\
        & full image & & 1.529 & 0.052 & $<$ 1.610 & \nodata & 0.77 & 0.12 & 1305 & 58\\
NGC 3626 & 15 \arcsec & 1.4 & 3.361 & 0.019 & 1.65 & 0.17 & 0.320 & 0.037 & 920 & 104\\
        & full image & & 5.578 & 0.084 & $<$ 2.546 & \nodata & 0.45 & 0.10 & 920 & 104\\
NGC 3665 & 15 \arcsec & 2.4 & 4.194 & 0.050 & 3.19 & 0.29 & 1.204 & 0.058 & 1570 & 26\\
        & full image & & 4.758 & 0.095 & \nodata & \nodata & 1.30 & 0.15 & 1570 & 26\\
NGC 4150 & 15 \arcsec & 1.0 & 1.205 & 0.015 & 0.599 & 0.086 & 0.136 & 0.043 & 1071 & 60\\
        & full image & & 1.474 & 0.064 & \nodata & \nodata & \nodata & \nodata & 1071 & 60\\
NGC 4429 & 15 \arcsec & 1.2 & 1.575 & 0.037 & $<$ 0.443 & \nodata & 0.493 & 0.022 & 1156 & 20\\
        & full image & & 1.83 & 0.11 & $<$ 1.485 & \nodata & 0.63 & 0.13 & 1156 & 20\\
NGC 4435 & 15 \arcsec & 1.2 & 1.808 & 0.044 & 0.46 & 0.14 & 0.878 & 0.050 & 1708 & 105\\
        & full image & & 1.71 & 0.13 & \nodata & \nodata & 0.95 & 0.14 & 1708 & 105\\
NGC 4459 & 15 \arcsec & 1.2 & 1.535 & 0.040 & $<$ 0.971 & \nodata & 0.339 & 0.017 & 1873 & 124\\
        & full image & & 1.85 & 0.10 & $<$ 2.469 & \nodata & \nodata & \nodata & 1873 & 124\\
NGC 4526 & 15 \arcsec & 1.2 & 7.465 & 0.053 & 2.24 & 0.28 & 4.813 & 0.054 & 6661 & 382\\
        & full image & & 8.690 & 0.092 & \nodata & \nodata & 4.822 & 0.078 & 6661 & 382\\
NGC 4694 & 15 \arcsec & 1.2 & 3.559 & 0.020 & 1.101 & 0.047 & 0.325 & 0.020 & 784 & 27\\
        & full image & & 4.435 & 0.049 & 1.36 & 0.15 & $<$ 0.341 & \nodata & 784 & 27\\
NGC 4710 & 15 \arcsec & 1.2 & 12.419 & 0.027 & 4.26 & 0.26 & 4.669 & 0.033 & 3726 & 26\\
        & full image & & 15.372 & 0.066 & 4.98 & 0.91 & 4.88 & 0.11 & 3726 & 26\\
NGC 5866 & 15 \arcsec & 1.1 & 5.75 & 0.16 & 1.58 & 0.17 & 0.53 & 0.15 & 7690 & 384\\
        & full image & & 14.11 & 0.94 & 3.98 & 0.83 & 0.49 & 0.15 & 7690 & 384\\
NGC 7465 & 15 \arcsec & 2.1 & 12.33 & 0.10 & 8.392 & 0.063 & 0.538 & 0.018 & 2804 & 33\\
        & full image & & 14.62 & 0.10 & 10.04 & 0.24 & 0.627 & 0.095 & 2804 & 33\\
UGC06176 & 15 \arcsec & 2.9 & 3.032 & 0.020 & \nodata & \nodata & 0.486 & 0.042 & 720 & 121\\
        & full image & & 3.332 & 0.067 & \nodata & \nodata & 0.44 & 0.12 & 720 & 121

\enddata
\tablecomments{Line fluxes and uncertainties for the PACS emission lines.  All quantities are in units of 10$^{-16}$ W m$^{-2}$.  Values for the full image less than 90\% of those obtained with a 15\arcsec\ are considered too noisy and are not shown.} %Brief caption (title)
\label{fluxtable}
\end{deluxetable*}

\setcounter{table}{1}
\begin{deluxetable}{ lcccc }
\tablecaption{SPIRE Line Fluxes and Uncertainties}
\tablehead{
 \colhead{Galaxy} & \colhead{Diam (\arcsec)} & \colhead{Flux} & \colhead{$\sigma$} & \colhead{Factor} } %Column headings
\tablecomments{Line fluxes and uncertainties for the SPIRE \nii 205 line in units of 10$^{-16}$ W m$^{-2}$. Because our sources are not point sources, the Semi Extended Correction Tool (SECT) was applied to our spectra, multiplying each channel by some scaling factor. The correction factor at 205 \micron\ is given.  To obtain an approximate value for the uncorrected flux, divide by the factor in the table.  \newline *NGC 2764 and 5866 have values listed for size estimates based on the 250 \micron\ data as well because the SECT corrected spectra produced from the CO estimates did not agree well with the photometry.}
\startdata

NGC 1222 & 12.5 & 0.607 & 0.018 & 1.257\\
NGC 1266 & 2.8 & 0.364 & 0.035 & 1.078\\
NGC 2764 & 6.4 & 0.679 & 0.016 & 1.060\\
NGC 3665 & 9.8 & 0.729 & 0.014 & 1.078\\
NGC 4459 & 9.0 & 0.309 & 0.010 & 1.079\\
NGC 4526 & 14.8 & 2.575 & 0.0048 & 1.061\\
NGC 4710 & 13.3 & 2.107 & 0.010 & 1.079\\
NGC 5866 & 10.1 & 0.579 & 0.0059 & 1.061\\
NGC 7465 & 12.7 & 0.333 & 0.024 & 1.078\\\tabularnewline\hline\tabularnewline
NGC 2764* & 27.3 & 1.168 & 0.027 & 1.812\\
NGC 5866* & 39.0   & 1.016 & 0.0087 & 1.818

\enddata
\label{SPIREdata}
\end{deluxetable}

%%%%---------------- Analysis --------------------------
\section{Analysis}
\label{sec:analysis}

Nearly every galaxy was detected in \cii, \oi, and \nii 122 $\mu$m lines as can be seen in Table \ref{fluxtable}. All nine galaxies observed by the SPIRE spectrometer were detected in the \nii 205\micron\ line.  To characterize the properties of the ISM in our selection of galaxies, the line versus continuum ratios are explored, as well as emission line ratios.  Additionally, \cii is investigated as both a SFR indicator and as a measure of H$_2$ mass. 

%Additionally, the ratio of \cii to FIR continuum flux depends on the heating efficiency of gas by photoelectrons off the dust grains, so the dependence of \cii/FIR on dust temperature (characterized by 60$\mu$m/100$\mu$m) is explored and compared to regular spiral galaxies.  Notably high values of \nii 122/\cii 158 are observed in Virgo Cluster members, and a general trend shows early type galaxies to have systematically low \oi 63/\cii 158 ratios.

\subsection{Line-to-Continuum Ratios}
\label{sec:continuum ratios}

The ratio of \cii to FIR continuum flux depends on the heating efficiency of gas by photoelectrons off the dust grains.  To explore such trends with dust temperature for the ETG sample, the three primary cooling lines \cii 158, \oi 63, and \nii 122 are plotted as a fraction of FIR flux against a proxy of dust temperature, the 60\micron/100\micron\ flux ratio. We calculate FIR flux using \iras\ 60 and 100 micron fluxes from \citet{Knapp}, and the formula given by \citet{Helou1988}. For NGC 3626 and UGC 06176, \iras\ data are unavailable, so AKARI 65 and 90 \micron\ fluxes are substituted \citep{Kokusho}.  For galaxies with Spitzer data (NGC 1266, 3032, 3489, 3607, 4150, 4435, 4459, 4526, and 5866), we calculate TIR flux using MIPS 24, 70, and 160 \micron\ data from \citet{Temi2} and the formula in \citet{DaleHelou}, then estimate FIR as TIR/2.  The FIR fluxes found from MIPS data are on average 79\% of those found with the IRAS data, ranging from 66\% to 96\% with better agreement found at lower dust temperatures.  The sum (\cii + \oi)/FIR is a measure of the total gas cooling efficiency and is shown alongside the other line/FIR plots in Figure \ref{linesFIR}. 

We make use of comparison samples obtained from \citet{Brauher} and the KINGFISH project \citep{Kenn2011}.  The galaxies in \citet{Brauher} contain all morphologies and were observed by \iso, while the KINGFISH galaxies are mostly late type galaxies and were observed with \herschel. Galaxies classified as ETGs from \citet{Brauher} that have velocities $>$ 3000 km s$^{-1}$ are really ULIRGS and are treated as LTGs in the comparison plots.  For line/FIR ratios for our sample galaxies we use the full image line flux values (unless they are less than the 15\arcsec\ fluxes) to more closely match the IRAS 60 and 100 \micron\ observations. For the KINGFISH galaxies, which are often extended and have good FIR images, we have computed FIR fluxes in apertures matching the line data (see Appendix \ref{sec:kingfish} for details).

The \cii/FIR values range from about 0.001 to 0.007, typical cooling efficiencies for the \cii line as can be seen by the overlap with other samples in Figure \ref{linesFIR}.  However, the line/continuum ratios are consistently lower than the KINGFISH sample.  Log values for our sample have a median value of -2.43 with a standard deviation of 0.22, while the KINGFISH LTGs have a median value of -2.10 with a standard deviation of 0.20.  This could be due to dust in ETGs being heated by older stars which won't release as many photoelectrons, resulting in less heating and a lower cooling efficiency.  We do not see a \cii/FIR deficit at warm dust temperatures (higher 60/100 \micron\ ratios) for ETGs, while the ratio decreases for the broader galaxy samples \citep{Brauher, Mal01}. In fact, a Kendall rank test indicates that \cii/FIR increases with dust temperature for our sample of ETGs, however the p-value for the rank test is 0.068, so the evidence for an increase with dust temperature is weak.

\citet{Smith2017} have also studied the \cii deficit in the KINGFISH galaxies with spatially resolved data, rather than single apertures as we use in our present work. They find that the trend of \cii/TIR increasing with 24 \micron\ surface brightness is more significant than the trend with IR color, and that the residuals from the 24 \micron\ trend also correlate with the gas metallicity. The values presented in Appendix \ref{sec:kingfish} reproduce those correlations.  Unfortunately, we do not have gas phase metallicities for the ETGs, so we cannot yet do a direct comparison on the role of gas metallicity in the ETGs. In any case it should be noted that \citet{Smith2017} demonstrate correlation, but not necessarily causation, since for the KINGFISH sample \cii/TIR and metallicity are also correlated with color, morphological type, and total stellar mass.  Thus additional exploration will be required before we can claim to know what drives the \cii deficits.

To begin some further exploration, we test an idea discussed by \citet{Mal01} and \citet{Smith2017}, which is that softer (redder) radiation fields have fewer photons available to ionize carbon and heat the gas. UV colors were taken from \citet{Young2014} (for ETGs) and \citet{Dale2007} (for KINGFISH galaxies).  The \cii/FIR ratio is strongly correlated with UV colors, shown in Figure \ref{UV_CIIFIR}. Our ETG sample yields a Kendall rank $\tau$ of -0.634 with a corresponding p value of 0.0001. Given this strong correlation, we also investigate the other line/FIR ratios with respect to UV color.

The \oi/FIR values range from roughly 0.0004 to 0.004.  For the most part \oi/FIR is constant with dust temperature, however there is evidence for the ratio to increase with dust temperature for ETGs.  A Kendall rank test on our sample returns a $\tau$ value of 0.45 and a corresponding probability of 0.004. The \oi/FIR ratios are on average slightly larger for the KINGFISH LTGs, with the log values having a median of -2.57 and standard deviation of 0.22, while our sample has a median of -2.73 and a standard deviation of 0.28. Like \cii/FIR, \oi/FIR also has a tendency to decrease with NUV-K colors, though not as sharply as \cii/FIR.  Kendall rank tests for both quantities in our sample yield $\tau$ values of -0.634 and -0.503 for \cii/FIR and \oi/FIR, respectively.    

The \nii 122/FIR values range from about 0.000125 to 0.00125, occupying a similar spread of values to comparison samples, with the log values having a median of -3.35 and standard deviation of 0.32, while the KINGFISH LTGs have a median value of -3.09 and a standard deviation of 0.18.  \citet{Mal01} and \citet{Brauher} find a trend of decreasing \nii/FIR with 60 $\mu$m/100 $\mu$m, and a similar gradual decline is present in our sample as well, however it is not statistically significant with a p value of 0.11.  In contrast to \cii/FIR and \oi/FIR, the \nii 122/FIR ratio has weak evidence of a positive trend with UV colors, with a $\tau$ of 0.281 and a p value of 0.052.  The ionization energy of nitrogen is 14.5 eV, so the \nii emission is more sensitive to the Lyman continuum fields of ionizing stars within \hii regions, whereas the \cii and \oi emission are more sensitive to the general interstellar radiation field reflected in the global galaxy UV-optical colors from GALEX photometry.

\citet{Mal01} find a slight decrease in gas cooling efficiency (\cii + \oi)/FIR with increasing dust temperature, however \citet{Brauher} claim to only find this decrease in spirals, but not in their sample of unresolved galaxies as a whole. A Kendall rank test shows a tendency for the efficiency to increase for ETGs. KINGFISH LTGs are consistently larger for this ratio as well, with log values having a median and standard deviation of -1.95 and 0.18, while our sample has a median and standard deviation of -2.20 and 0.23. As with the \cii/FIR and \oi/FIR ratios, the total cooling efficiency decreases with redder UV-optical colors.

\subsection{Emission Line Ratios}
\label{sec: line ratios}

\subsubsection{\oi/\cii as a Gas Density Tracer in ETGs}
\label{sec: oi to cii ratios}

Carbon has an ionization potential of 11.26 eV, while the ionization potential of oxygen is 13.62 eV, so in PDR layers where H is neutral and atomic, carbon is C+ and oxygen is neutral.  The \oi line has a higher critical density and excitation energy than \cii, so it is the primary coolant of warm, dense gas, and in x-ray dissociated regions, the \oi line can be stronger than in typical PDRs \citep{Maloney}. As their FIR lines have different critical densities, the line ratio \oi/\cii can be used as a density indicator.  This is complicated by the fact that in deeper layers of the PDR carbon is neutral (or in CO) alongside atomic oxygen.  Also C+ exists in ionized gas while neutral O does not.  Another complication is the fact that the \oi line can be optically thick for some values of hydrogen density, $n$, and interstellar radiation field strength, $G_0$. Despite these complications, the \oi/\cii ratio can be used as a number density probe for dense gas (3 $\times$ 10$^3$ < $n$ < 5 $\times$ 10$^5$ cm$^{-3}$).  For $n$ significantly less than the critical density of an optically thin emission line, assuming a fixed column density, the intensity increases linearly with $n$.  For $n$ significantly greater than the critical density, the intensity becomes independent of $n$. The critical density for \cii is about 3 $\times$ 10$^3$ cm$^{-3}$ and the critical density for \oi is 5 $\times$ 10$^5$ cm$^{-3}$, so the \oi/\cii ratio should increase throughout this range. 

\setcounter{figure}{4}
\begin{figure}
\epsscale{1.1 } %Override for scale, in decimal units, e.g., 0.80
\plotone{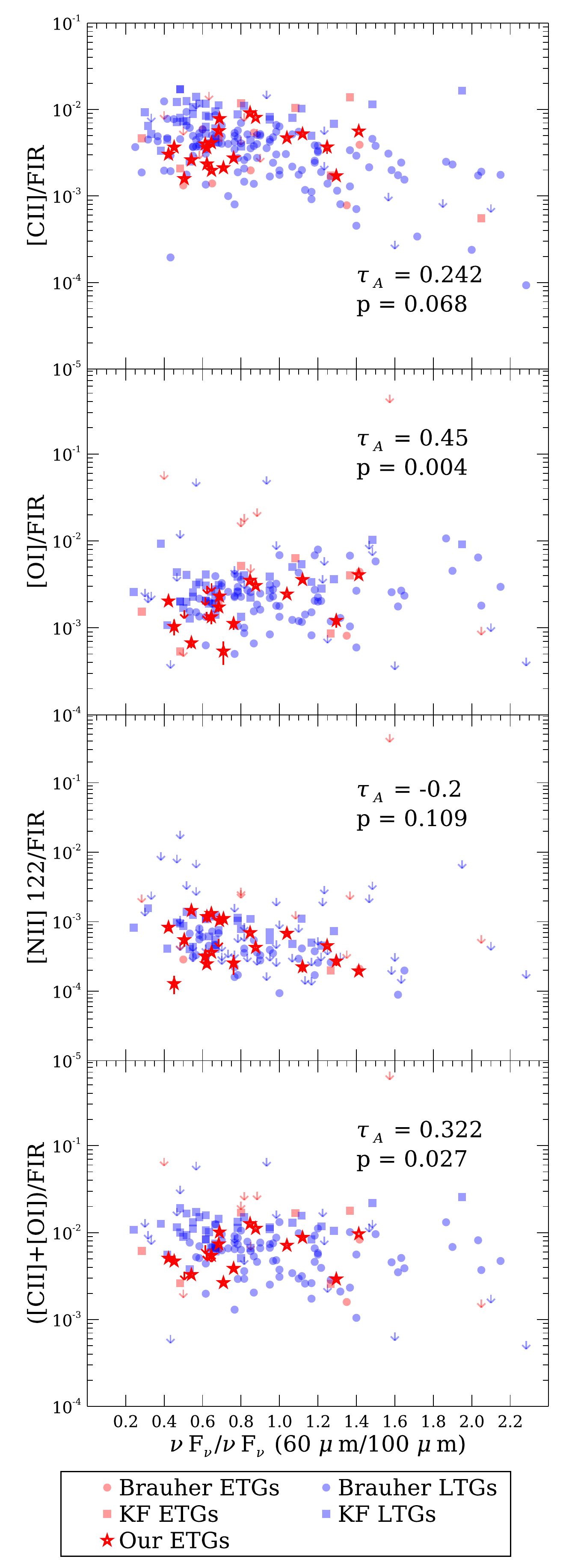} %Plots one file, image width scaled to text width
\caption{Each PACS line flux relative to FIR plotted against a tracer of dust temperature. The red stars are our sample, the blue circles are LTGs from \citet{Brauher}, and the blue squares are LTGs from the KINGFISH survey \citep{Kenn2011}. Red circles and squares are ETGs from the corresponding samples.  Kendall rank $\tau$ coefficients and their corresponding p values are also shown.  Full image flux values are used for our sample unless the 15\arcsec\ flux is greater. If data from our sample is missing error bars, the error is  smaller than the symbol.}
\label{linesFIR}
\end{figure}

\begin{figure}
\epsscale{1.1}
\plotone{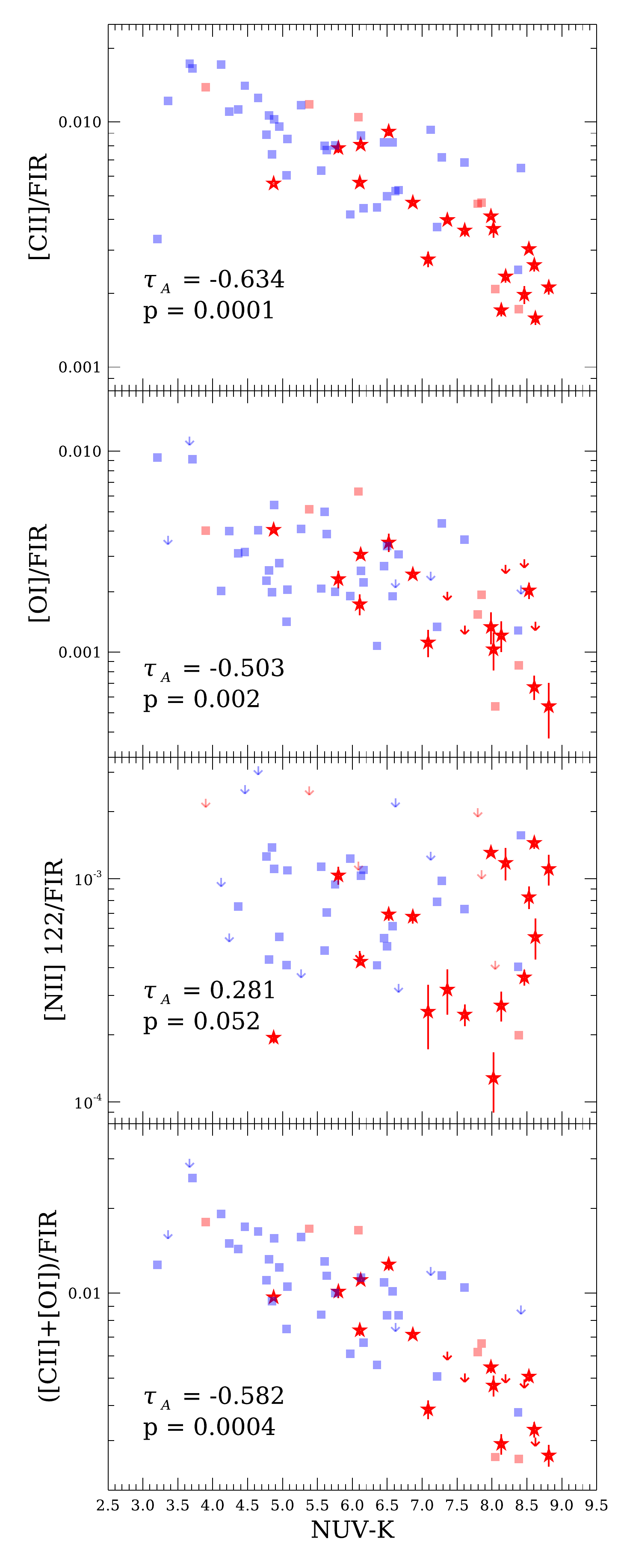}
\caption{The line/FIR ratios plotted against NUV-K.  KINGFISH LTGs are plotted as blue squares, while KINGFISH ETGs are plotted in red. Our ETGs are plotted as red stars. For each line Kendall rank $\tau$ coefficients and their corresponding p values are also shown. For our ETGs there is a strong anti-correlation over several magnitudes for \cii/FIR and NUV-K, which seems to drive the correlation of \nii/\cii with UV colors. There does not appear to be a correlation of \nii/FIR with UV colors. If data from our sample is missing error bars, the error is  smaller than the symbol.}
\label{UV_CIIFIR}
\end{figure}

\begin{figure}
\epsscale{1.2 } %Override for scale, in decimal units, e.g., 0.80
\plotone{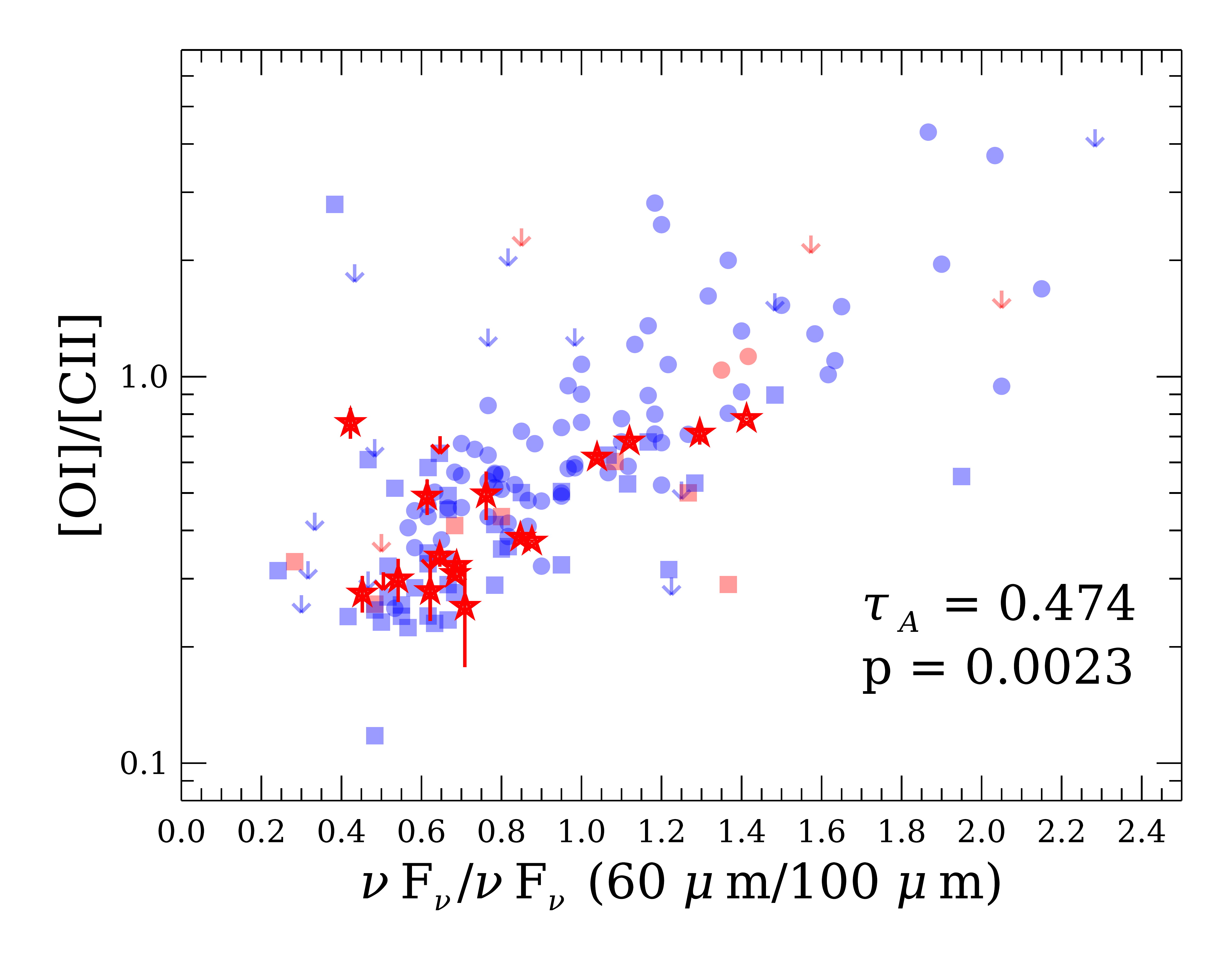} %Plots one file, image width scaled to text width
\caption{The \oi 63 to \cii 158 line ratio, plotted against a tracer of dust temperature. The red stars are our sample (from the 15\arcsec\ apertures), blue and red circles are LTGs and ETGs, respectively, from the \citet{Brauher} sample, and the blue and red squares are LTGs and ETGs, respectively, from the KINGFISH study \citep{Kenn2011}.  Our sample has on average lower \oi/\cii line ratios than the Brauher sample. If data from our sample is missing error bars, the error is  smaller than the symbol.}
\label{OCratio}
\end{figure}

For the line/line ratios we use the 15\arcsec\ aperture fluxes since those apertures define exactly the same region on the sky in both lines. The line ratio \oi/\cii ranges from 0.25 to 0.78, consistent with regular spirals and other ETGs, but lower than AGN and ULIRGs, which can approach values as high as 5 \citep{Brauher}.  The median of the log values from \citet{Brauher} is -0.19 with a standard deviation of 0.25, but the KINGFISH LTGs have a median closer to the value from our sample.  The median from KINGFISH is -0.48 with a standard deviation of 0.23, while the median from our ETGs is -0.46 with a standard deviation of 0.21. There is a general trend for this ratio to increase with increasing dust temperature, and a Kendall Rank test on our sample returns a $\tau$ = 0.474 and a corresponding p = 0.0023. Thus the ETGs exhibit the same behavior as normal spirals in this respect.  These results can be seen in Figure \ref{OCratio}.

%good line for PDR models section%
%Assuming that the \oi/\cii is signaling gas density and that the Schmidt law for star formation holds, we expect a higher star formation efficiency in regions with higher \oi/\cii ratios.  A high SFE results in a higher $G_{0}/n$ and thus higher dust temperatures.

\subsubsection{Large \nii/\cii Ratios in ETGs }
\label{sec: nii to cii ratios}

The \nii emission lines arise exclusively in regions where hydrogen is ionized since the ionization potential of nitrogen is 14.5 eV.  Collisions with electrons are the excitation mechanism for these emission lines, so the \nii 122 and 205 \micron\ lines can be used to infer electron density.  The lines are also used to estimate the \cii emission arising from ionized gas. Both \cii and \nii are used as star formation indicators (e.g. \citet{HCamus, Zhao2016}), and our results below suggest that for some kinds of samples there may be a systematic bias between the two measures.  Understanding what drives the systematic difference between them is crucial if they are to be widely used.  Looking for correlations with other parameters will help clarify their relationship.  

\begin{figure*}
\epsscale{1.2 } %Override for scale, in decimal units, e.g., 0.80
\plotone{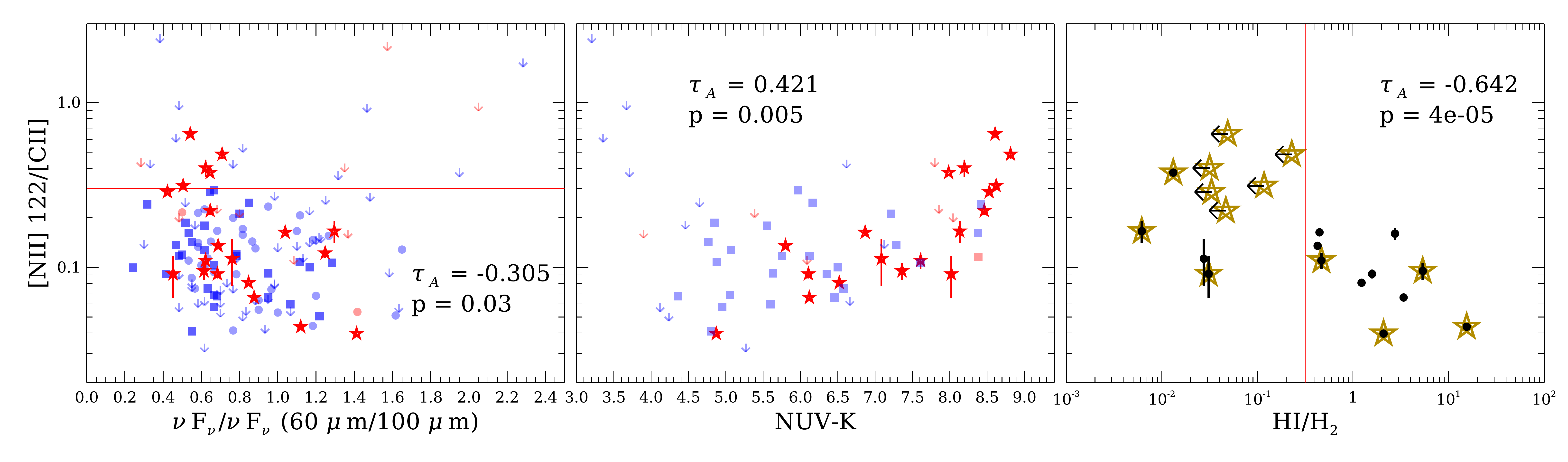} %Plots one file, image width scaled to text width
\caption{The \nii 122 to \cii 158 line ratio, plotted against a tracer of dust temperature, UV colors, and the atomic to molecular gas mass ratio. Our sample has large \nii/\cii ratios for several galaxies. For the first two panels, ETGs are red, while LTGs are blue.  Our sample is plotted as stars, while the KINGFISH sample is plotted as squares \citep{Kenn2011}, and the Brauher sample is plotted as circles \citep{Brauher}. In the first panel a red line is drawn at \nii/\cii = 0.3, above which there are no detections other than those from our ETG sample. The second panel shows a correlation with UV colors. For ETGs there is a linear correlation with a Pearson r value of 0.70.  A two tailed test for the p-value yields 0.0012. The last panel shows that the highest \nii/\cii values have Log(\hi/H$_2$)$<-0.5$.  \hi values were obtained from \citet{Serra}, while H$_2$ values came from \citet{Young2011}.  Galaxies with AGN are indicated with gold stars in panel 3. If data from our sample is missing error bars, the error is  smaller than the symbol.}
\label{NCratio}
\end{figure*}

The line ratio \nii 122/\cii has values from 0.035 to 0.64, and is plotted against several different quantities in Figure \ref{NCratio}.  We find that 25\% of our sample have ratios significantly larger than comparison spirals and ETGs, which seem to reside primarily below a value of 0.3 (drawn in the first panel of the figure).  There are a couple of galaxies from the KINGFISH sample that approach these high ratios, including several upper limits, but no detections are as high as ours, the largest of which is a factor of 2 larger than any previous measurement.  The large line ratios reside exclusively at low dust temperatures as seen in the first panel of Figure \ref{NCratio}.  \citet{Cormier} find small \nii 122/\cii ratios for low metallicity dwarf galaxies, however without gas phase metallicity measurements we cannot test whether the inverse is true. According to the models of \citet{Mittal}, ratios as high as the ones we observe can be produced if there are only small contributions to \cii from neutral gas (i.e. PDRs). These high line ratios may be related to the lower atomic gas fraction in their host galaxies, and indeed we only find the larger \nii 122/\cii ratios in galaxies with Log(\hi/H$_2$)$<-0.5$, shown in the third panel of Figure \ref{NCratio}.

Another possible influence is the presence of an AGN in galaxies with higher \nii/\cii ratios, indicated by stars in the third panel of Figure \ref{NCratio}.  It has been proposed that AGN activity decreases the \cii emission by increasing the ionization state of the gas  (i.e. carbon will mostly be C2+ rather than C+) \citep{Mal01}.  All galaxies with a ratio $>0.2$ (NGC 3607, NGC 3665, NGC 4429, NGC 4435, NGC 4459 NGC 4526, and NGC 4710) are presumed to have an AGN \citep{Kristina}. However, some galaxies with an AGN have small \nii/\cii ratios (i.e. NGC 1222 with a ratio of 0.040). Those without an AGN have ratios between 0.066 and 0.14 and those with an AGN range between 0.035 and 0.64.  An AGN may be a necessary condition, but not a sufficient condition for an elevated \nii/\cii ratio. 

Even more striking is evidence that the hardness of the general interstellar radiation field, as measured by UV-optical colors, are strongly correlated with the \nii/\cii ratio.  UV colors were used to plot the \nii/\cii ratio against NUV-K and an obvious trend can be seen in panel two of Figure \ref{NCratio}. For the ETGs, a Kendall rank coefficient of 0.421 was found for \nii/\cii and NUV-K.  A two tailed test for p-values yields 0.005, showing that the correlation is highly unlikely to arise by chance. The ratio is related to the strong correlation of \cii/FIR with UV colors shown in Figure \ref{UV_CIIFIR}, and corresponding lack of correlation for \nii/FIR, caused by the difference in ionization potentials for carbon and nitrogen.

%so the \nii emission may be more sensitive to the Lyman continuum fields of ionizing stars within \hii regions, whereas the \cii emission may be more sensitive to the general interstellar radiation field reflected in the global galaxy UV-optical colors from GALEX photometry.

Among all the emission lines in the ETG sample, there do not appear to be any fundamental differences to help differentiate cluster galaxies from field galaxies.  Both line/FIR ratios and line/line ratios span the same range for high or low values of $\rho_{10}$ and $\Sigma_3$ (both indicators of galaxy density) as defined in \citet{Paper7}.  Additionally we find no obvious relation between the ratios and the importance of organized rotation versus non-ordered velocity dispersion (as defined by $\lambda_{Re}$ in \citet{Paper3}), or any relation to the misalignment of stars and gas \citep{Paper10}.

Further analysis of the line ratios will be explored in a future paper.  A detailed analysis of the gas properties using PDR models will be done, utilizing the \ci and high J CO lines from the SPIRE observations in conjunction with the PACS emission lines and ancillary data. 

\subsection{Fraction of \cii Arising in the Ionized Gas of ETGs}
\label{sec:ionized}

Since the ionization potential of carbon is lower than hydrogen, carbon will be ionized in \hii\ regions, but to complicate matters ionized carbon can be found in PDRs as well, alongside neutral carbon and CO near the dense center of the gas cloud.  To properly use PDR models and calculate line ratios, these two components must be decomposed into a \cii flux attributed to the ionized region and a component attributed to PDRs. We first tried to use the method in \citet{Mal01} to separate these components, but in several cases this attributed all of the \cii emission to ionized regions, which is unlikely.

\begin{figure}
\epsscale{1.2 } %Override for scale, in decimal units, e.g., 0.80
\plotone{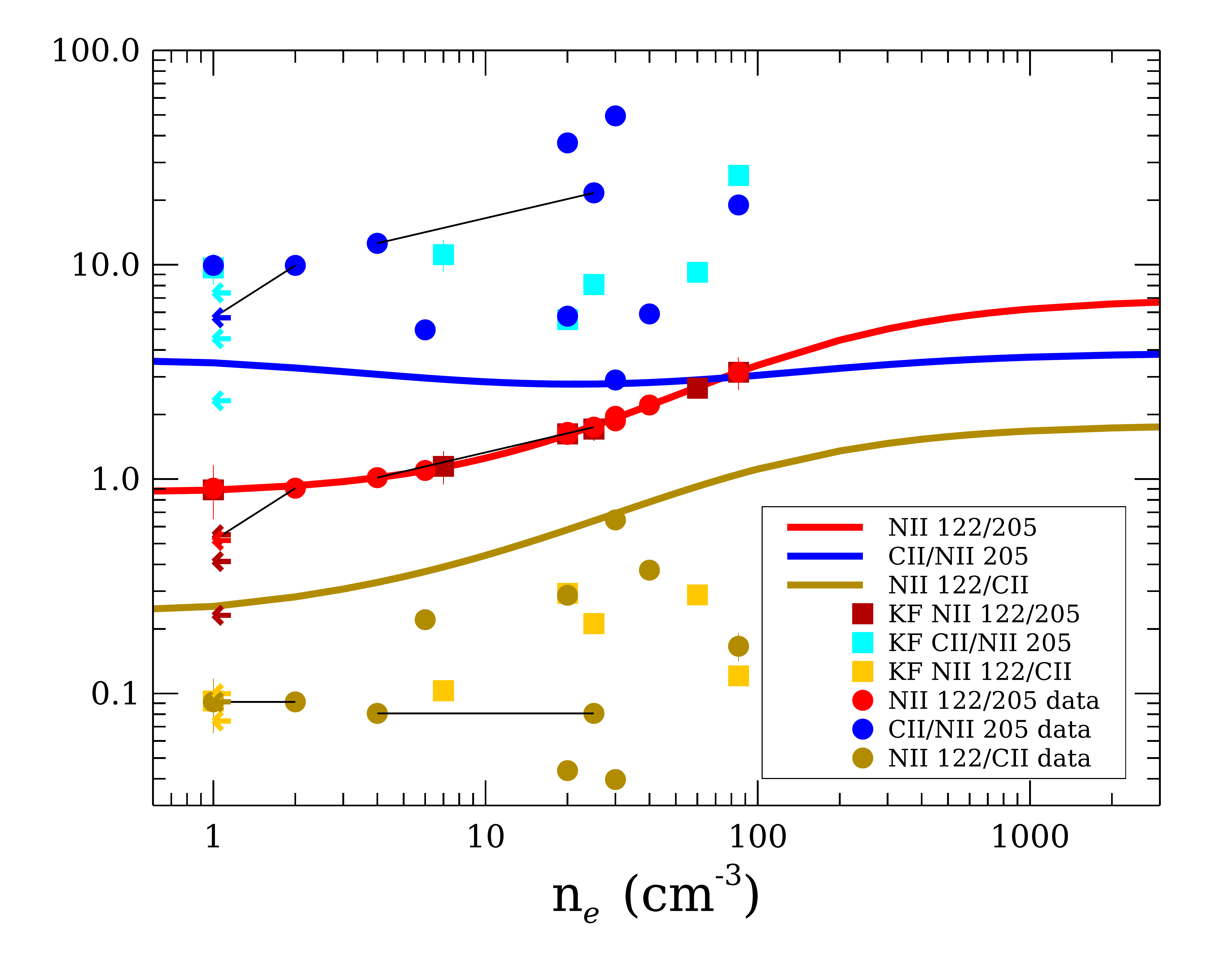} %Plots one file, image width scaled to text width
\caption{Expected line ratios from collisional excitation models.  A two level atomic model was used for \cii and a three level model was used for \nii. Data for NGC 2764 and 5866 have two points connected by a line for data obtained with both CO and 250 \micron\ size estimates.}
\label{NitroRatios}
\end{figure}

A detailed method for decomposing the \cii flux into neutral and ionized components is discussed in \citet{Oberst} for galaxies with observations of the \nii 122 and 205 lines. Figure \ref{NitroRatios} depicts this method at work for both our sample galaxies (circles) and the KINGFISH sample galaxies (squares).  Assuming electron impact as the excitation method, the electron density of the ionized regions can be found using the line ratio of \nii 122/\nii 205 and the computed red line.  This electron density can then be used to calculate an expected line ratio for  \cii/\nii 205 for ionized gas, according to the blue line.  This ratio along with the \nii 205 flux predicts the ionized \cii flux,  which may be subtracted from the total to find the \cii flux attributable to PDRs.  Note the same method may be performed using the \nii 122/\cii ratio instead (yellow line), but this ratio is more sensitive to uncertainties in the density. The results of decoupling the \cii emission into ionized and neutral components are in Table \ref{CIIiontable}. Upper and lower limits for the contribution from PDRs were calculated by using the upper and lower limits on the \nii 122/\nii 205 ratio. Our ETGs and the KINGFISH galaxies tend to have similar \cii$_{PDR}$ components.  Our early type galaxies average 63.5\% (70.9\% without NGC 4526), while the KINGFISH galaxies average 53.0\% (59.6\% without NGC 5457). A comparison like this was not possible prior to \herschel observations of the \nii 205 line.  These results depend on the assumed \nii/\cii abundances and further emission line data will be necessary to check those assumptions.

Theoretical line ratios were calculated using a three level atom model for \nii and a two level atom model for \cii.  These assumptions are adequate because the third energy level for \nii has E/k of approximately 22000 K, while the second energy level for \cii has E/k of approximately 62000 K.  The relative abundances were drawn from \citet{Savage} and corrected for ionization states based on models from \citet{Rubin}.  The gas phase abundances used were n(C+)/n$_e$ = 1.4$\cdot 10^{-4}$ \citep{Cardelli} and n(N+)/n$_e$ = 7.9$\cdot 10^{-5}$ \citep{Meyer}.  Collision strengths were scaled to an assumed electron temperature of 8000 K, and were taken from \citet{Hudson} for nitrogen and \citet{Blum} for carbon.  Values were computed with updated collision strengths from \citet{TayalCII} and \citet{TayalNII} and differ by 5-10\%.  

There seem to be more galaxies with large contributions to \cii emission from ionized gas than predicted by models \citep{Accurso}. Table \ref{CIIiontable} shows that four (or possibly five) galaxies from our sample as well as four KINGFISH galaxies have roughly half or more than half (47.9\% to 100\%) of their \cii emission from ionized gas.  For galaxies without \nii 205 measurements, we can use the prescription of \citet{Mal01} that found that more than half the \cii emission of a galaxy is from ionized gas if its \nii 122/\cii $\geq$ 0.12.  This criterion would concern IC 0676, NGC 3032, NGC 3607, NGC 4429, NGC 4435, and UGC 06176, with NGC 3489 and NGC 4150 both having ratios greater than 0.11 as well. In addition to these galaxies, the results of \citet{Parkin} show M51 to have 80\% of its \cii emission coming from ionized gas in the center, with declining values at larger radii. For IC 342, \citet{Rollig} find that in the central few hundred pc, 70\% of the \cii emission comes from ionized gas.  We do not find such a high value using the KINGFISH data, and this is due to a discrepancy in the \cii/\nii205 ratio. They adopt a value of 4.89, while we use 2.90, and this factor of 2 is passed on to the ionized gas contribution.

Attributing these relatively high proportions of \cii emission to ionized gas may have implications for high redshift galaxies and galaxy evolution.  Theoretical models of high z galaxies often ignore the contribution to \cii from ionized gas, or if they calculate it, they find that it is small. \citet{Popping16} do not account for \cii emission from ionized gas, while \citet{Olsen} find the contribution from ionized gas to be under 3\%. \citet{Vallini} also find a very small contribution to the \cii flux from ionized gas, and assume an electron density much less than the critical density of n$_e$ = 8 cm$^{-3}$, while we infer electron densities as high as 85 cm$^{-3}$, with over half of our galaxies having densities greater than 8 cm$^{-3}$. Previously most local galaxies had confirmed small contributions to the \cii flux from ionized gas, but now larger samples are finding cases that prove this assumption false. We should be careful to understand local galaxies in order to ensure the high redshift models are correct.

NGC 4526 is an extreme case and warrants further consideration.  The galaxy is not unusual in other respects (i.e. \cii/FIR, \oi/\cii, \cii/CO, etc.), yet the adopted N/C abundance and \nii/\cii ratios suggest that 96\% of \cii emission arises from ionized gas instead of PDRs. It is reminiscent of NGC 4125, where \citet{Wilson} propose that all of the \cii emission comes from ionized gas, but there is one major difference between the two cases:  in NGC 4125 the assertion about ionized gas rests heavily on the fact that neither cold atomic gas (\hi), molecular gas (CO), nor \oi have been detected.  In NGC 4526 we detect abundant molecular gas, so it's curious that PDR emission apparently plays such a small role.  The cause could be optical depth effects in the \cii line (e.g. \citet{Langer}), combined with a soft ISRF (very red colors), high gas density in the PDR (so that the C+ shell is physically thin, with a low column density), and/or unusual N/C abundance (as discussed for intermediate mass AGB stars in \citet{Mittal}).

%Where to put this sentence?
%
%When decomposed into PDR and ionized gas components, \cii$_{PDR}$/FIR tends to decrease while \cii$_{ion}$/FIR increases with UV color. 

\setcounter{table}{2}
\begin{deluxetable}{ lccccl }
\tablecaption{Separating \cii Emission Components}
\tablehead{
 \colhead{Galaxy} & \colhead{$\frac{[NII]122}{[NII]205}$} & \colhead{$\frac{[NII]122}{[NII]205} \sigma$} & \colhead{e$^{-}$} & \colhead{$\frac{[CII]}{[NII]205}$} & \colhead{ion\%}  } %Column headings
\tablecomments{The results of decoupling ionized and neutral components of \cii using the \nii 122/\nii 205 ratio for our sample as well as the KINGFISH sample \citep{Kenn2011}.  Electron densities predicted by the model are listed (cm$^{-3}$), as well as the expected \cii/\nii 205 ratio.  The amount of \cii emission arising from ionized regions can then be inferred from the \cii/\nii 205 ratio. The percent of \cii emission coming from ionized gas is given in the last column. \newline *NGC 2764 and 5866 have additional entries listed for size estimates based on the 250 \micron\ data as well as the CO sizes because the SECT corrected spectra produced from CO sizes underestimate the photometry, while the 250 \micron\ sizes overestimate the photometry.  The true values for these galaxies likely lie in between.}

\startdata

NGC 1222 & 1.96 & 0.084 & 31 & 2.79 & 5.6$_{-0.0}^{+0.0003}$ \\
NGC 1266 & 3.15 & 0.56  & 85 & 3.00 & 15.8$_{-0.006}^{+0.003}$\\
NGC 2764 & 1.75 & 0.067 & 24 & 2.77 & 12.8$_{-0.0001}^{+0.0}$\\
NGC 3665 & 1.65 & 0.086 & 21 & 2.77 & 48.2$_{-0.0}^{+0.0006}$\\
NGC 4459 & 1.10 & 0.066 & 6  & 2.96 & 59.6$_{-0.01}^{+0.02}$\\
NGC 4526 & 1.87 & 0.021 & 28 & 2.78 & 96.1$_{-0.004}^{+0.0}$\\
NGC 4710 & 2.22 & 0.019 & 40 & 2.82 & 47.9$_{-0.0}^{+0.0}$\\
NGC 5866 & 0.915 & 0.26 & 2 & 3.30 & 35.1$_{-0.06}^{+0.03}$\\
NGC 7465 & 1.62 & 0.13 & 20 & 2.77 & 7.5$_{-0.0}^{+0.0003}$\\
NGC 2764* & 1.02 & 0.038 & 4 & 3.08 & 24.5$_{-0.005}^{+0.007}$\\
NGC 5866* & 0.522 & 0.15 & $<1$ & 3.78 & 66.2$_{-0.0}^{+0.0}$\\\tabularnewline\hline\tabularnewline
IC 0342 & 2.66 & 0.23 & 60 & 2.90 & 31.5$_{-0.004}^{+0.004}$ \\
NGC 1097 & 1.63 & 0.05 & 20 & 2.77 & 50.0$_{-0.0}^{+0.0}$ \\
NGC 2146 & 3.15 & 0.30 & 85 & 3.00 & 11.5$_{-0.002}^{+0.002}$ \\
NGC 3521 & 0.55 & 0.15 & $<1$ & 3.75 & 50.7$_{-0.0}^{+0.0}$ \\
NGC 4826 & 1.71 & 0.21 & 25 & 2.78 & 34.3$_{-0.0}^{+0.001}$ \\
NGC 5055 & 0.41 & 0.14 & $<1$ & 3.75 & 82.9$_{-0.0}^{+0.0}$ \\
NGC 5457 & 0.23 & 0.06 & $<1$ & 3.75 & 100$_{-0.0}^{+0.0}$ \\
NGC 5713 & 0.89 & 0.15 & 1 & 3.49 & 36.1$_{-0.05}^{+0.03}$ \\
NGC 6946 & 1.15 & 0.20 & 7 & 2.92 & 26.2$_{-0.01}^{+0.03}$

\enddata
\label{CIIiontable}
\end{deluxetable}

\subsection{\cii as a Star Formation Tracer}
\label{sec:CIISFR}

\begin{figure}
\epsscale{1.2 } %Override for scale, in decimal units, e.g., 0.80
\plotone{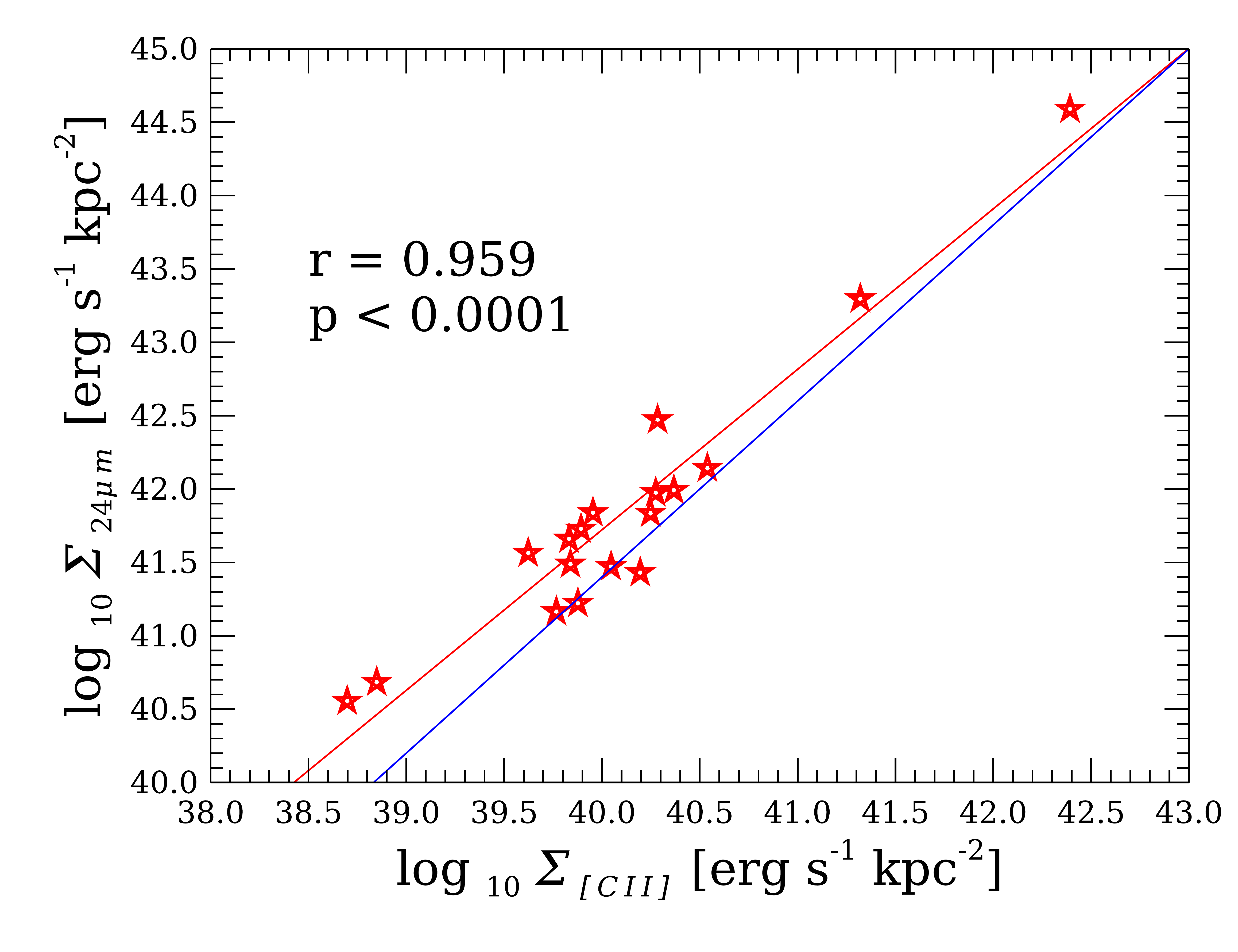} %Plots one file, image width scaled to text width
\caption{24 $\mu$m surface brightness (corrected for circumstellar contribution) plotted against \cii surface brightness for the galaxies in our sample. The line of best fit for our sample is shown in red, while the line of best fit found by \citet{HCamus} is shown in blue.}
\label{CIISFRfig}
\end{figure}

The \cii line is a major coolant of interstellar gas, and the main source of heating for the gas in spirals is FUV photons from O and B stars, so it is not surprising that strong correlations between the \cii surface brightness and several star formation tracers have been found by \citet{Boselli}, and more recently studied by \citet{HCamus}. To compare to their work, we obtained Spitzer MIPS 24 $\mu$m fluxes for our galaxies from NED, or IRAS 25 $\mu$m when 24 $\mu$m was unavailable (this was the case for IC 0676, IC 1024, NGC 1222, NGC 2764, NGC 3665, NGC 4429, NGC 4710, and NGC 7465) \citep{Knapp,Sanders,Moshir,Temi,Temi2}.  There were no data available for NGC 3626 or UGC 06176. If the reported flux was not corrected for the circumstellar contribution, the correction at 22 \micron\ used by \citet{Davis} was applied.  Most corrections were only a few percent, except for NGC 3665 and NGC 4429 which had 60\% and 35\% of the flux removed, respectively. Additional data from the WISE 22 micron survey were available for all galaxies in the sample, and were utilized as a consistent, similar comparison.  The WISE data, along with estimated surface areas for each galaxy (used for both 22 and 24 \micron), were taken from \citet{Davis}. As in Figure 2 of \citet{HCamus}, we plot the 24 $\mu$m surface brightness against the \cii surface brightness (including the contribution from ionized gas) and find a similar result.  There is a tight correlation with an r-value of 0.959 and a standard deviation of 0.27 dex.  Our data are plotted with the best fit lines in Figure \ref{CIISFRfig}. The slope we find is 1.094 $\pm$ 0.080 (red line) which is slightly less than the value of 1.2 (blue line) found by \citet{HCamus}, but quite close. The standard deviation on their data was 0.23 dex, so \cii is an equally good star formation tracer for ETGs as LTGs, although the conversion equation is slightly different. It should be noted that our values are galactic values while \citet{HCamus} use multiple regions from each galaxy to form their sample.  The results indicate that \cii is a useful SFR probe in all types of galaxies, and could be particularly useful in high z galaxies where traditional methods may be unavailable.

\subsection{\cii as a Molecular Gas Tracer}
\label{sec:CIICO}

\begin{figure}
\epsscale{1.2}
\plotone{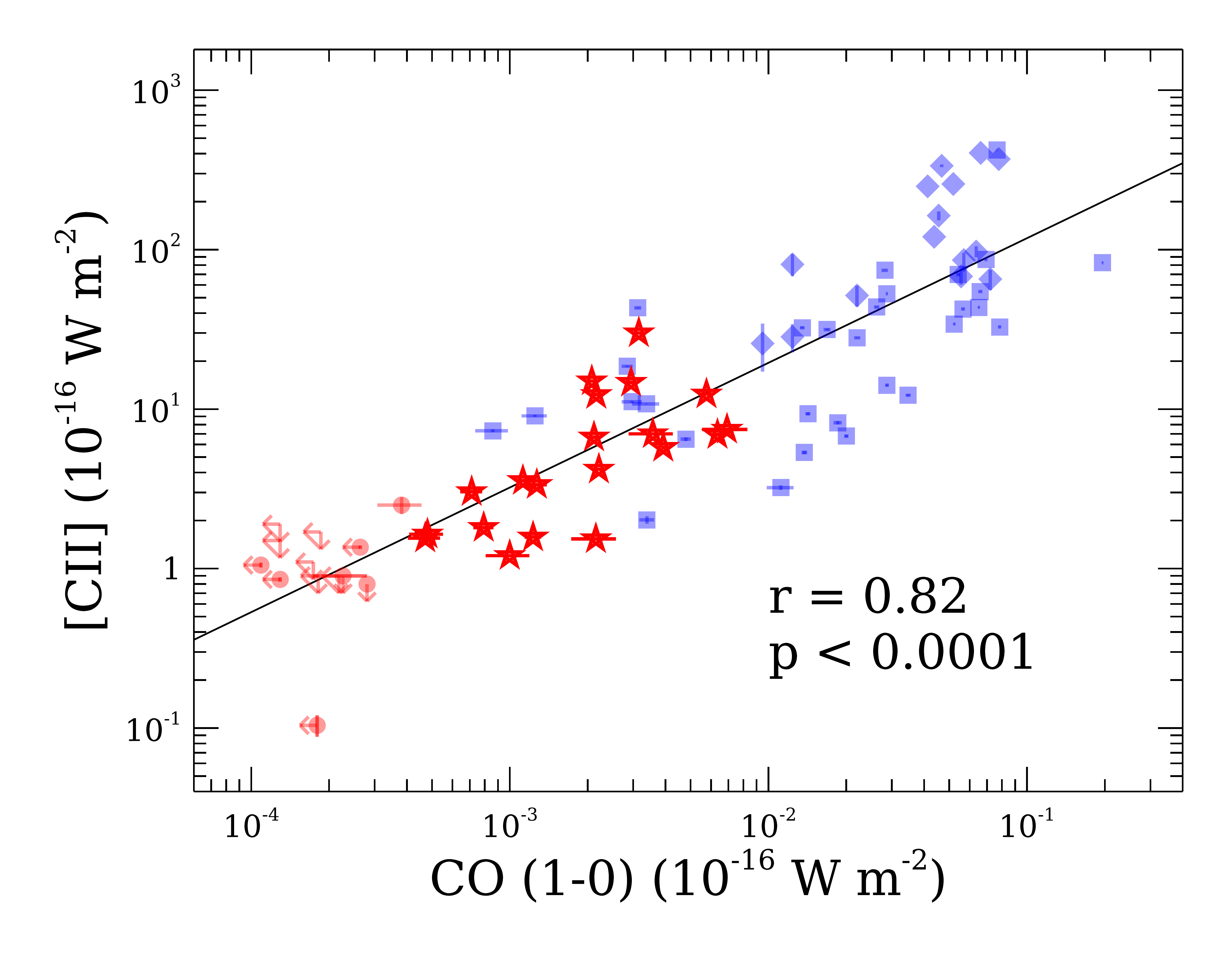}
\caption{\cii emission vs CO (1-0) emission.  Red stars are from our sample, while blue diamonds come from \citet{Stacey}, blue squares are LTGs from the KINGFISH and HERACLES projects (Appendix A; \citep{Leroy2009}), and red circles are CO poor ETGs from \citet{Brauher} and \citet{Werner}. If data from our sample is missing error bars, the error is  smaller than the symbol.}
\label{CIIvsCOfig}
\end{figure}

Observations of $^{12}$CO(1-0) were reported for the majority of galaxies in our sample in \citet{Young2011}, with a few coming from \citet{Young2008}, an observation for NGC 3489 coming from \citet{Crocker2011}, NGC 3607 from \citet{Welch}, and upper limits for NGC 4374, 4486, 5813, and 5846 from \citet{Combes} and NGC 4636 from \citet{Sage}.  While \cii can originate in many different phases of the interstellar medium, the majority of the \cii emission is presumed to originate from the surface of molecular clouds in PDRs (except for a few galaxies with large \cii$_{ion}$ contributions), typically around 70\% as shown earlier. Because \cii is much brighter than CO, it could be used as a proxy for molecular gas in galaxies where CO emission is faint such as high redshift sources.  CO resides at the center of the clouds where \cii originates, so there should be CO emission correlated with the \cii emission.  As shown in Figure \ref{CIIvsCOfig}, this is indeed the case, and we find our results in good agreement with those obtained by \citet{Stacey}.  The trend is also consistent with additional CO poor galaxies from the Atlas3D sample that also have \cii data available (NGC 4278, 4374, 4461, 4477, 4486, 4503, 4596, 4608, and 5322 \citep{Brauher} and NGC 4472, 4636, 5813, and 5846 \citep{Werner}).   The data are plotted on a log-log scale in Figure \ref{CIIvsCOfig}. The log-log slope is 0.78 $\pm$ 0.068, with a Pearson coefficient of r=0.82.  This linear relation in log space implies a power law conversion between \cii and CO, with \cii (W m$^{-2}$) $=$ 714$\cdot$CO(1-0)$^{0.78}$ (W m$^{-2}$). The results also show that \cii has a similar scatter with CO in ETGs as LTGs. The \cii/CO ratio can also be used along with other ratios as an indicator of the PDR gas density and UV field strength through the use of PDR models as in \citet{Gullberg}.  A full study of PDR models will be carried out in a subsequent paper, but we note that similar \cii/CO ratios in ETGs and LTGs suggest we will find similar gas densities and UV fields.

Though CO is the preferred method for quantifying the molecular gas content in galaxies, it has been shown that \cii can trace "CO dark" molecular gas that resides in the layers of PDRs that are not traced by CO.  \citet{Langer2014} find that molecular clouds in the Milky Way contain significant amounts of CO dark molecular gas, from 20\% in dense clouds to 75\% in diffuse clouds.  This CO dark gas can be a large fraction (up to 80\%) of the total molecular gas mass in low metallicity galaxies such as dwarf galaxies \citep{Fahrion}.  We do not expect such large amounts of CO dark gas in our galaxies with roughly solar metallicity, but several galaxies in our sample could contain an appreciable amount of CO dark molecular gas.  To estimate the amount of molecular gas traced by \cii, we calculate the H$_2$ column density using equation 4 from \citet{Langer2014} and a relative abundance X$_{H_{2}}$(C+) = 2.8$\cdot$10$^{-4}$, assuming a pressure of 10$^4$ K cm$^{-3}$.  We do not correct our \cii fluxes for contributions from atomic or ionized gas, so the mass estimate is an upper limit. On average, $\leq$ 17\% of the total molecular mass in our galaxies is CO dark.  The galaxies that contain the most CO dark gas are NGC 7465, IC 1024, and NGC 3032, being comprised of $\leq$ 46\%, 39\%, and 35\% CO dark gas, respectively.  The H$_2$ column densities and masses can be seen in Table \ref{COdarkgas}.

\begin{deluxetable}{ lccccc }
\tablecaption{CO dark gas}
\tablehead{
 \colhead{Galaxy} & \colhead{H$_2$ } & \colhead{Area } & \colhead{CO dark gas} & \colhead{CO gas} & \colhead{\% of total gas}  } %Column headings
\tablecomments{We have calculated upper limits on the amount of CO dark gas in the galaxies in our sample.  Listed here are the H$_2$ column densities (10$^{21}$ cm$^{-2}$), the galaxy areas (kpc$^2$) \citep{Davis}, the masses found from \cii and CO (log(M$_\odot$)), and the percent of the total molecular mass that is CO dark.}

\startdata

IC 0676&1.71&2.51&7.84&8.63&13.9\\
IC 1024&3.85&4.29&8.42&8.61&39.3\\
NGC 1222&7.71&1.63&8.3&9.07&14.6\\
NGC 1266&1.78&0.03&5.93&9.28&0\\
NGC 2764&3.77&8.63&8.72&9.19&25.1\\
NGC 3032&1.8&4.85&8.14&8.41&35.1\\
NGC 3489&0.42&0.5&6.53&7.2&17.5\\
NGC 3607&0.4&8.14&7.71&8.42&16.5\\
NGC 3626&0.86&1.51&7.32&8.21&11.4\\
NGC 3665&1.08&8.84&8.18&8.91&15.8\\
NGC 4150&0.31&1.32&6.81&7.82&9\\
NGC 4429&0.4&0.98&6.8&8.05&5.3\\
NGC 4435&0.46&0.57&6.63&7.87&5.4\\
NGC 4459&0.39&1.91&7.08&8.24&6.5\\
NGC 4526&1.92&2.22&7.83&8.59&14.9\\
NGC 4694&0.91&1&7.16&8.01&12.5\\
NGC 4710&3.19&2.97&8.18&8.72&22.4\\
NGC 5866&1.48&2.39&7.75&8.47&16\\
NGC 7465&3.16&10.42&8.72&8.79&46.1\\
UGC 06176&0.78&1.87&7.37&8.58&5.8

\enddata
\label{COdarkgas}
\end{deluxetable}

\FloatBarrier

%%%%---------------- Conclusion --------------------------
\section{Conclusion}
\label{sec:conclusion}

There are very few observations of FIR emission lines from ETGs, while many similar observations have been carried out for other morphological types, including spirals, dwarf galaxies and (U)LIRGs. Here we present \herschel observations of 20 ETGs from the Atlas3D survey, including \cii, \oi, and \nii 122 emission lines observed by PACS and \nii 205 emission observed by SPIRE for a subset of 9 the galaxies. 

For the most part we find the empirical measurements of the ETGs occupy the same ranges as other types of galaxies.  Line to FIR ratios are consistent with previous observations, as well as most of the \oi/\cii and \nii 122/\cii ratios.  The \oi/\cii ratios do not reach values as high as ULIRGs or AGN, but are similar to regular spiral galaxies. Twenty-five percent of our \nii 122/\cii ratios are higher than any other galaxies, though some galaxies from the KINGFISH survey have large \nii 122/\cii upper limits.  There is some evidence to suggest this may be correlated with the state of the ISM as indicated by the atomic/molecular gas ratio, and it is certainly correlated with the UV colors. Better knowledge of the elemental abundance in these galaxies would help disentangle these correlations. 

The \cii/FIR ratio is strongly anti-correlated with NUV-K and likely gives rise to the correlation of \nii 122/\cii with UV colors. Redder NUV-K colors are indicative of a soft ISRF and would be relatively weak in the UV photons that ionize carbon or cause photoelectric heating of the gas. However, the \nii 122 emission suggests that there are non-negligible amounts of harder ($>$ 13.6 eV) UV radiation.  Overall it appears the global gas properties in ETGs are quite similar to those in late type galaxies.

We find that \cii is useful as a star formation tracer in ETGs. The scatter found for ETGs is similar to normal spirals, which provides indirect support for their use in high redshift galaxies, as the \cii line has been shown to be a star formation tracer for all morphological types of local galaxies. Molecular gas surface density is also known to be a good predictor of star formation rates, and since \cii originates on the surface of clouds harboring CO, it naturally traces $^{12}$CO(1-0) emission very well, too, and we calculate a power law conversion of \cii (W m$^{-2}$) $=$ 714$\cdot$CO(1-0)$^{0.78}$ (W m$^{-2}$).  

The percent of \cii emission arising in ionized gas is similar for ETGs and normal spirals, except for a few unusually high values in NGC 3665, NGC 4459, NGC 4526, NGC 4710, and possibly NGC 5866, which reach values similar to those seen in the center of M51 and in NGC 4125. About a quarter of galaxies from our sample, as well as the KINGFISH survey, have more than half of their \cii emission coming from ionized gas, contrary to the current theoretical models for high redshift galaxies, and could lead to overestimates when \cii is used to trace CO-dark gas in PDRs. This discrepancy could have significant implications for the future study of high redshift galaxies and galaxy evolution, where \cii is quickly becoming a popular diagnostic workhorse.

Further study of ETGs using FIR spectroscopy will be carried out using the SPIRE observations of 9 galaxies from our present sample.  The pair of neutral carbon emission lines in the SPIRE spectra will allow us to better characterize the neutral atomic gas in PDRs, constraining the pressure at the boundary between atomic and molecular gas and providing information about the density and radiation field strength through their use in PDR models.  Neutral carbon can also trace another layer of "CO dark" molecular gas that may be present in PDRs. Additionally, the mid to high J CO emission lines found in the SPIRE spectra will help characterize the excitation of the molecular gas in these galaxies, as well as constrain the density at the center of molecular clouds.

Metallicity is also a very important variable that has not been incorporated into our study of ETGs.  Gas phase metallicity measurements will be used to explore trends in all of the FIR line ratios, but the \cii/FIR and \nii 122/\cii will be of particular interest.

%%%%---------------- Acknowledgements --------------------------
\acknowledgments{{\it Acknowledgments:}  

This work is based on observations made with Herschel, a European Space Agency Cornerstone Mission with significant participation by NASA. Support for this work was provided by NASA through an award issued by JPL/Caltech.

%Recognition of facilities used. Optional listing of specific instrument in parentheses: \facility{HST(WFPC2)}
Facilities: \facility{\herschel (PACS)}, \facility{\herschel (SPIRE)}

\setcounter{figure}{1}
\begin{figure*}
\epsscale{ 2} %Override for scale, in decimal units, e.g., 0.80
\includegraphics[width=\textwidth,height=\textheight,keepaspectratio]{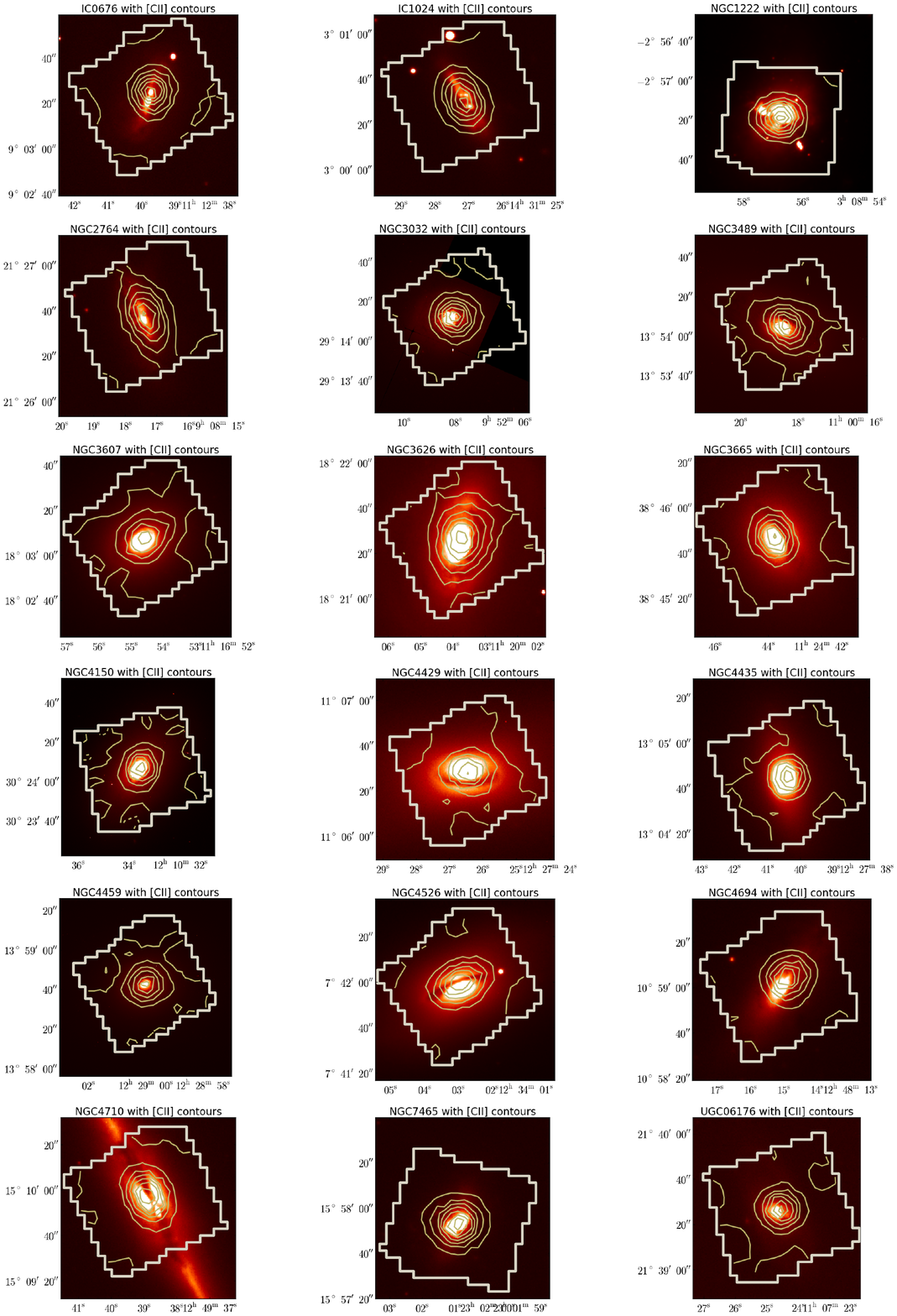}
\caption{ Optical images of the galaxies are seen along with the PACS field of view and \cii 158 contours around each galaxy. }
\label{optical}
\end{figure*}
  
\begin{figure*}
\epsscale{2} %Override for scale, in decimal units, e.g., 0.80
\includegraphics[width=\textwidth,height=\textheight,keepaspectratio, trim = 0mm 40mm 0mm 0mm]{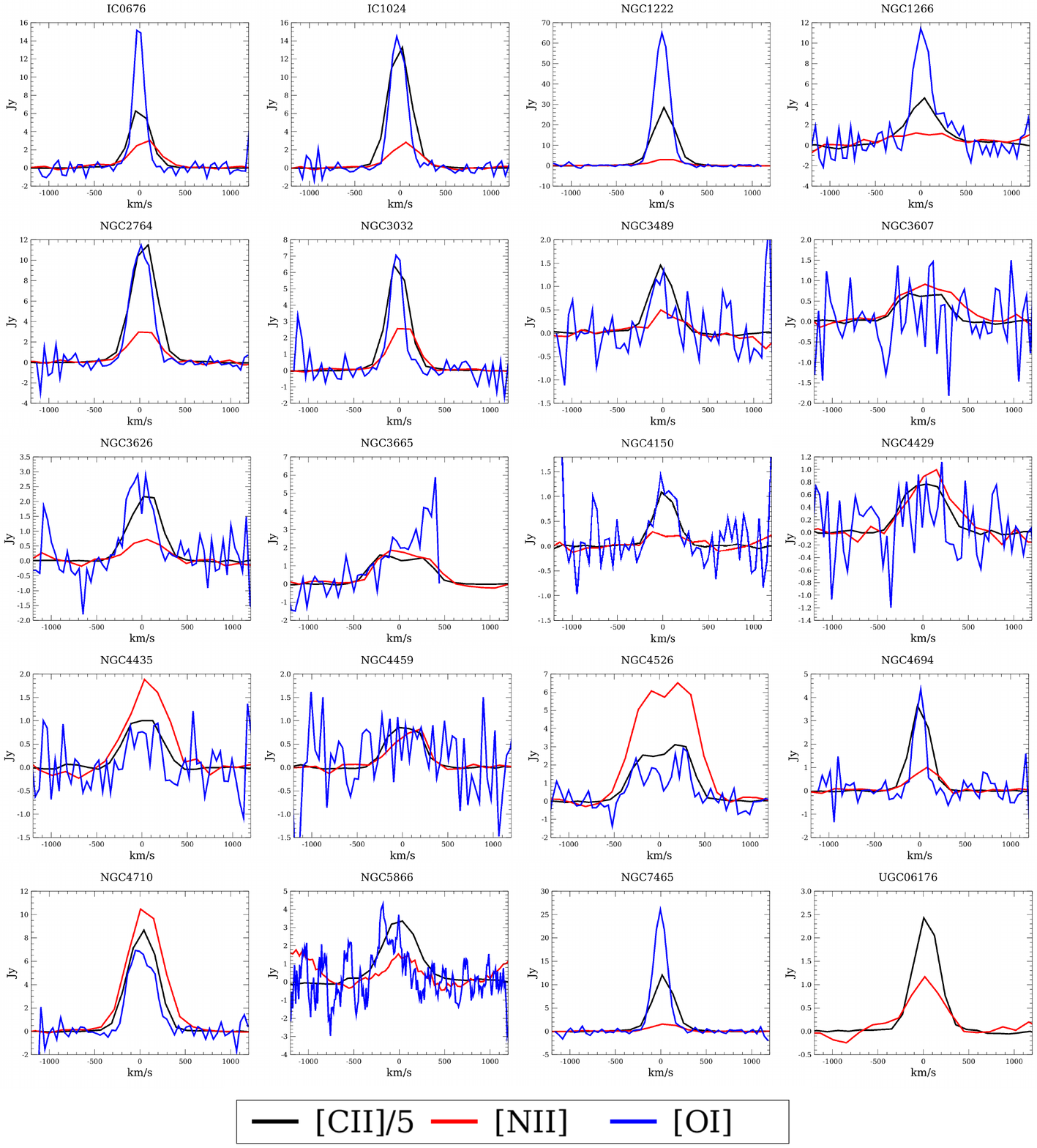} 
\caption{ The \cii 158, \oi 63, and \nii 122 emission spectra are shown in black, blue, and red respectively.  The \cii flux is scaled down by a factor of 5. The spectra shown were obtained by spatially integrating within a 15\arcsec\ aperture. The spectral resolutions for the \cii, \nii, and \oi lines are 237.8 km/s, 291.2 km/s, and 87.5 km/s, respectively.}
\label{spectra}
\end{figure*}

\begin{figure*}
\epsscale{2}
\includegraphics[width=\textwidth,height=\textheight,keepaspectratio, trim = 0mm 80mm 0mm 10mm]{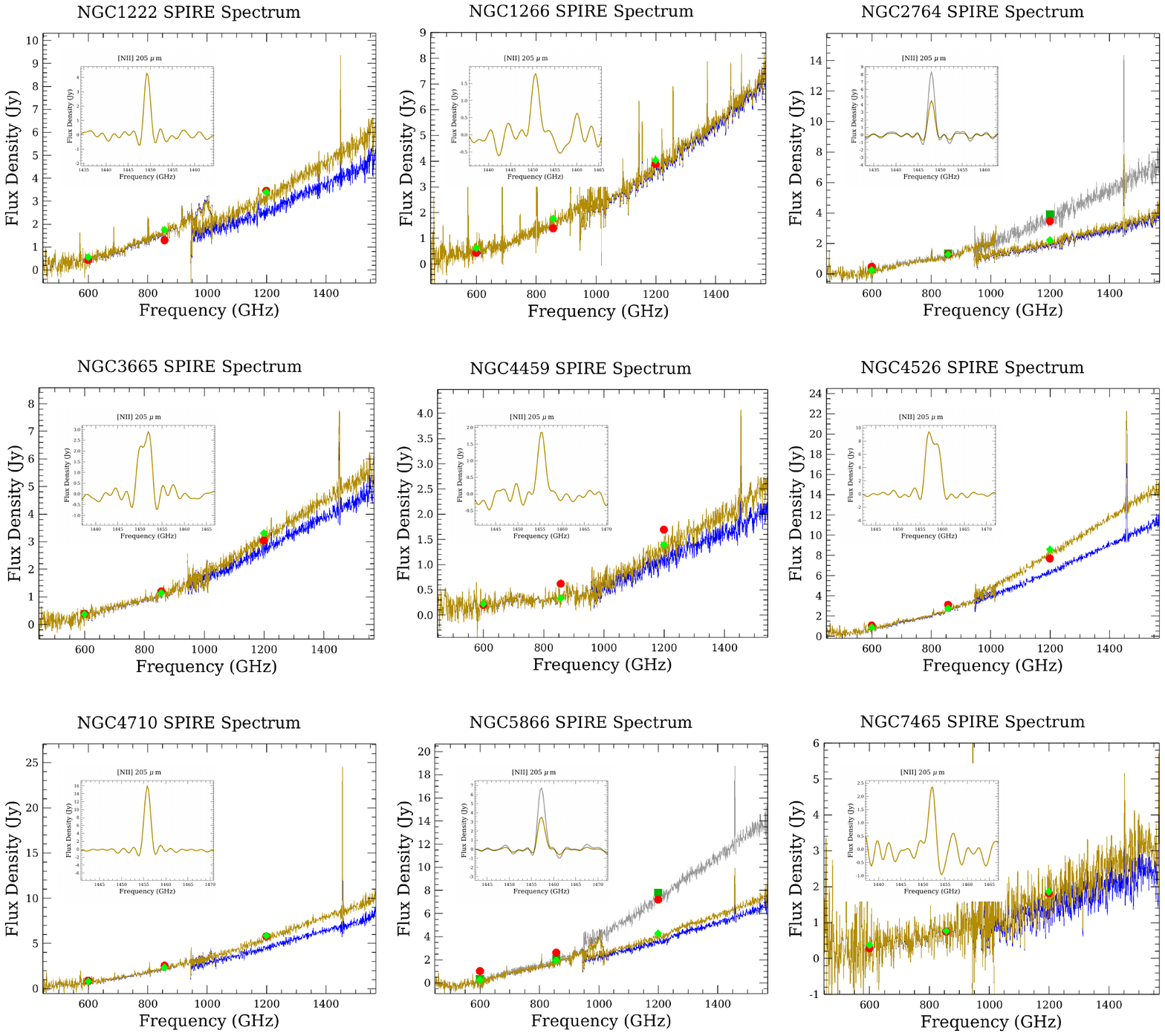}
\caption{ The SPIRE spectra for every galaxy.  The uncorrected spectra are blue, while the spectra corrected by the SECT are shown in gold for source sizes based on CO images, and in gray for source sizes based on 250 \micron\ images (only NGC 2764 and 5866). Overlaid in red are photometric data points obtained with 22, 30, and 42 \arcsec\ aperture radii for the 250, 350, and 500 $\micron$ SPIRE photometry bands, respectively, except for NGC 2764 and 5866, where a 28 \arcsec\ radius was used for the 250 \micron\ band. The aperture radii were chosen to contain just the main lobe of the beam, but for NGC 2764 and 5866 a larger aperture was necessary to contain the extended emission. The green diamonds are synthetic photometry data produced by the spireSynthPhotometry task with the CO size corrected spectra, and the darker green squares are synthetic photometry based on the 250  \micron\ size corrected spectra. }
\label{SPIREspectra}
\end{figure*}

\FloatBarrier

%%%%---------------- Appendix --------------------------

\clearpage
\begin{appendix}
%\appendix
\section{KINGFISH Data}
\label{sec:kingfish}

FIR fine structure line flux measurements are available for a large sample of nearby galaxies, mostly spirals, in the KINGFISH project \citep{Kenn2011}.  In this paper we make use of the PACS spectroscopic data for the \cii, \nii 122\micron, \oi 63\micron, and \nii 205\micron\ lines. Here we present some specific notes related to the measurement of those line fluxes, comparisons to previous measurements, and notes on corresponding measurements of TIR and CO fluxes.

\subsection{PACS Lines}

\subsubsection{Data Processing}

Fundamentally the line fluxes in the KINGFISH galaxies are measured in the same manner as for our sample of early-type galaxies (Section \ref{sec:lineflux}), with the following exception.  In estimating the line flux uncertainties, we note that the expression $\sqrt{N} \sigma_1 \delta v$ is only appropriate when the individual channels are statistically independent of each other.  However, the projected cubes in the Herschel Science Archive are produced with the parameter upsample = 4, meaning that the output channels are 4 times narrower than Nyquist sampling and the noise in adjacent channels is strongly correlated. Direct experimentation with this parameter suggests that, when upsample = 4, the expression above needs to be multiplied by an extra factor of 2 to produce an accurate uncertainty estimate.

The choice of aperture sizes and locations in the KINGFISH data also deserves some comment. The KINGFISH galaxies tend to be closer than our early-type galaxies, so to achieve broad similarity in linear aperture sizes between the two samples we usually adopt apertures of radius 20\arcsec\ for the KINGFISH galaxies.  We take a minimum linear size of 500 pc for the nearest galaxies, and in two cases the aperture size is enlarged by a factor of 3 because the line emission is in a ring around the nucleus. Figure \ref{fig:apertures} shows the distribution of aperture sizes. The KINGFISH apertures are usually centered on the galaxy nucleus, but in nine cases the aperture is off the nucleus because the \nii field does not cover the nucleus or the \nii emission is stronger elsewhere.  For example, in NGC4736 the aperture we use for the PACS lines is centered not on the nucleus but on the eastern side of the ring (see \citet{van2015} for images).  Detailed information on the aperture positions and sizes is in Tables \ref{KFtable} and \ref{KFtable2}. 

\subsubsection{Comparison to Previously Published Values}
\label{sec:kfcompare}
Our measured fluxes and line ratios are consistent with previously published values, though it is notable that there have been some significant calibration updates in the most recent release of the HIPE software.\footnote{http://herschel.esac.esa.int/twiki/bin/view/Public/HipeWhatsNew14x} Specifically, (1) a new correction for the illumination pattern on the spaxels is advertised to increase the PACS line fluxes by up to $\sim$ 30\% compared to previous calibrations; and (2) the \nii 205\micron\ lines are fully reprocessed with a new relative spectral response function (RSRF) that accounts for the red leak at wavelengths longer than 190\micron. Thus, we provide direct comparisons of older line fluxes to new ones measured in the KINGFISH data reprocessed to SPG v14.2.0.

The KINGFISH Data Release 3 (KFDR3), released in October 2013, is available from the Herschel Science Center and was based on HIPE v11.  We find that line fluxes in the newly reprocessed SPG v14.2.0 data are systematically higher than in KFDR3 by about 35\% for \oi 63\micron\ and about 20\% for \nii 122\micron\ and \cii.  (The \nii 205\micron\ data are discussed separately below.)  Thus, to the accuracy available in these tests, line ratios like \nii122/\cii do not change between KFDR3 and SPG v14.2 but the \oi63/\cii line ratios are higher in v14.2 than in KFDR3 by a median value of 11\%. Furthermore, line ratios to external quantities (e.g. \cii/FIR and \cii/CO) increase by $\sim$20\% over previous calibrations.  Recent analyses such as \citet{HCamus, HCamus2016, HCamus2017} and \citet{Smith2017} appear to use the KFDR3 calibrations. The v14.2 values are used in this paper and quoted in the tables. 

Older published values are roughly consistent with the values presented here.  \citet{Crox2012} plot \oi/\cii ratios for many individual 10\arcsec\ apertures in NGC1097 and NGC4559; in NGC4559 their ratios fall in the range of 0.2 to 0.33, and our measured value is $0.276 \pm 0.008$.  In NGC1097 they find a similar range, though bright nuclear regions are much closer to 0.33; our value, which is strongly dominated by the nuclear ring, is $0.453 \pm 0.003$.  In the ring of NGC4736, \citet{van2015} quote ratios from 0.17 to 0.33, and  we measure $0.326 \pm 0.011$.

For the PACS \nii 205\micron\ line, the new RSRF effectively multiplies fluxes up by factors of 5 to 10 compared to previous versions including KFDR3. The recalibrated fluxes are also in good agreement with measurements from the SPIRE FTS \citep{Kam2016}.  Specifically, there are six \nii 205\micron\ detections in Table \ref{KFtable} which also have roughly coincident pointings with SPIRE and the mean flux ratio (SPIRE/PACS) of those seven is 0.63, with a dispersion of 0.10.  The agreement is remarkable given that the aperture shapes and sizes are not particularly well matched; the SPIRE beam at 205\micron\ is roughly a Gaussian of FWHM 17\arcsec, and \citet{Kam2016} have attempted to extrapolate to a 43.5\arcsec\ beam, whereas for the PACS data we have used uniform circular apertures whose radius is usually 20\arcsec. The \nii 122/205 line ratios calculated from Table \ref{KFtable} are also roughly consistent with analogous values obtained by \citet{HCamus2016} using the older (KFDR3) PACS 122\micron\ data and SPIRE FTS 205\micron\ fluxes. Since those authors do not publish tables, only the rough envelopes of measured values can be compared, but the range of their values ($\approx$ 0.6 to 6, with a peak around 1.4) is similar to ours ($\approx$ 0.5 to 4, also with a peak between 1 and 2). 

\subsection{TIR and CO data}

Table \ref{KFtable} also provides FIR fluxes for the relevant apertures in the KINGFISH galaxies, for use in calculating \cii/FIR and similar ratios over those apertures.  The ``total infrared'' (TIR) fluxes are estimated from the 100\micron\ PACS images in KFDR3 using calibration coefficients in Table 2 of \citet{Galametz}; those authors also show that for the KINGFISH galaxies, the 100\micron\ emission can be accurately scaled to TIR emission with very little dependence on IR color.  For comparison with ``far-IR" (FIR) fluxes derived from IRAS 60\micron\ and 100\micron\ flux densities, we note that TIR is equal to FIR$\times 2$, within a few percent, for a wide range of IR colors \citep{DaleHelou}. \citet{Galametz} estimate that the calibration uncertainty in the PACS images is about 7\%, and the dispersion about their 100\micron -TIR relation is 0.05 dex (12\%).

CO imaging data for many of the KINGFISH galaxies are also available as observations of the CO J=2$-$1 line from the HERACLES project \citep{Leroy2009}, which used the IRAM 30m telescope in an on-the-fly mapping mode.  Those data have an intrinsic spatial resolution of 12\arcsec\ and are modestly smoothed to produce a spatial resolution of 13.4\arcsec\ FWHM, similar to the 11.5\arcsec\ FWHM of the PACS \cii data.  Again, we measure line fluxes by constructing circular apertures within which the data are spatially integrated to make a single spectrum.  For the purposes of analyzing \cii/CO ratios and their spatial variation, the apertures are centered on the galaxy nucleus and have radii varying from 13.4\arcsec\ to 300\arcsec.  Since the \cii fields of view often consist of a roughly square area or a long rectangular strip, the CO data cubes are masked to match the \cii fields prior to spatial integration.  The CO spectra are straightforward to integrate, with one exception.  The CO data cube for NGC4559 also seems to show, besides the target galaxy in velocities 700 to 900 km s$^{-1}$, additional significant emission at a velocity of about 1180 km s$^{-1}$.  This emission has a  20\% effect on the CO fluxes for the galaxy, and it is counted in the \cii/CO line ratio because the corresponding velocity range is unavoidably included in the \cii line.  

In comparing the KINGFISH \cii/CO(2$-$1) line ratios and ours, which are derived from CO(1$-$0), it is necessary to assume a CO (J=2$-$1)/(J=1$-$0) line ratio.  For the galaxies in the HERACLES project, \citet{Leroy2009} find a mean brightness temperature ratio $T_B(J=2-1)/T_B(J=1-0) = 0.8$, meaning that integrated line luminosity or flux ratios will be 6.4 in energy units.  Similarly \citet{Stacey2010} have used a ratio of 7.2 for their sample galaxies, which are primarily ULIRGS and may have somewhat higher CO excitation.  \citet{Leroy2009} assume an absolute calibration uncertainty of 20\% for the HERACLES data, based on repeated observations of some bright objects on multiple days; they also remind readers that this value does not account for relatively poor knowledge of the aperture efficiencies.

Table \ref{KFtable2} gives \cii and CO J=2$-$1 fluxes and statistical uncertainties, and the \cii/CO(2$-$1) ratio, for all of the galaxies with both KINGFISH and HERACLES data.  The exceptions are Ho II, IC2574, NGC3190, NGC4594, and NGC5474, where the CO emission is faint.  The table quotes the line fluxes and ratio measured in the largest aperture, whose radius is given in columns 2 and 3; variations in in this global \cii/CO ratio from one galaxy to another span a factor of nearly 20, which is certainly larger than both the statistical uncertainties and the calibration uncertainties.  

\subsection{Radial Gradients in Line Ratios}

These datasets also provide information on the radial gradients in line ratios, and because the KINGFISH apertures tend to have smaller physical sizes than the apertures on our early-type galaxies (Figure \ref{fig:apertures}), it is useful to try to quantify the effects of aperture size.  \citet{Smith2017} discuss some extreme cases in which nuclear \cii/TIR ratios are a factor of 2 to 5 lower than values measured at $\ge$ 1 kpc, but in most cases the radial gradients in \cii/TIR are modest.  Similarly, we have computed \cii/TIR ratios for the KINGFISH galaxies in a nuclear aperture of radius 15\arcsec\ and in the largest aperture available; the ratios systematically increase with aperture size, but the median increase is only 0.09 dex, which is much smaller than the factor of 10 range exhibited by the set.  Figure \ref{fig:apertures} also shows, as an example, that the trend of \cii/TIR with global galaxy color is robust to the aperture size distribution. Thus, line ratio gradients are certainly present in these datasets but they do not dominate the observed trends.

We also find that the \oi/\cii and \nii122/\cii ratios are usually constant with radius in the KINGFISH galaxies.  There are a few notable cases (especially NGC3627, NGC4736, and NGC6946) where \oi/\cii increases sharply, by a factor of a few, for points within about 1 kpc of the nucleus.  In IC0342, NGC1097, NGC4321, NGC5457, and NGC6946 we find similarly dramatic increases in \nii122/\cii in the nucleus. The coverage of the \nii fields often does not extend farther than 1 kpc, though, so these discussions of radial variations are necessarily incomplete. Quantitative analysis of the line ratios is outside the scope of this paper, but we note that a trend of higher \oi/\cii line ratios in galaxy nuclei would be roughly consistent with higher gas densities and UV field strengths, as expected (e.g. \citet{Mal01}; \citet{van2015}, and many others).  The KINGFISH galaxies with nuclear jumps in \oi/\cii and \nii/\cii also tend to have AGN (see, e.g. \citet{HCamus} or \citet{Smith2017}).

Most of the galaxies in this sample show a general trend with lower \cii/CO ratios in their nuclei (inner 0.5 to 1 kpc) and higher values at larger radii.  In an extreme case like NGC4736 the \cii/CO surface brightness ratio varies from a low of $130 \pm 10$ in the nucleus (radii $< 300$ pc) to a high of $1300 \pm 300$ in the ring (1 to 1.5 kpc), and a moderate $\approx 650$ beyond; however, in most cases the gradients are not that steep.  Generally speaking the lower \cii/CO line ratios in galaxy nuclei are qualitatively consistent with an expectation for higher gas densities in nuclei (e.g. \citet{Stacey2010, Gullberg}).  There does not seem to be a simple or obvious relationship between the presence or absence of an AGN and a radial trend in the \cii/CO ratio.

\FloatBarrier
\clearpage
\begin{turnpage}
\setcounter{table}{0}
\renewcommand{\thetable}{A\arabic{table}}
\begin{deluxetable*}{ lcccccccccccccc}
\tablecaption{KINGFISH Line Fluxes and Uncertainties}
\tablecomments{Columns 2 and 3 give the center position of the chosen aperture.  Column 4 gives the distance from this aperture center to the nucleus of the galaxy in the plane of the sky.  Columns 5 and 6 give the radius of the aperture.  Subsequent columns give the integrated line fluxes within those apertures, with their 1 $\sigma$ statistical uncertainties or 3 $\sigma$ upper limits.  The quoted uncertainties include only the effects of noise in the spectra.}
%Data from the KINGFISH sample.  

\tablehead{
 \colhead{Galaxy} & \colhead{RA} & \colhead{ Dec} & \colhead{Offset} & \colhead{radius} & \colhead{radius} & \colhead{\cii} & \colhead{\cii$\sigma$} & \colhead{\oi} & \colhead{\oi$\sigma$} & \colhead{\nii122} & \colhead{\nii122$\sigma$} & \colhead{\nii205} & \colhead{\nii205$\sigma$} & \colhead{FIR}\\
\colhead{} & \colhead{} & \colhead{} & \colhead{(kpc)} & \colhead{(kpc)} & \colhead{(\arcsec)} & \multicolumn{9}{c}{(all fluxes and uncertainties in units of 10$^{-16}$ W m$^{-2}$)}} 

\startdata

   HOII &124.80817&70.71811&0.67&0.5&33.8&2.22&0.22& $<$ 1.38 & \nodata  & $<$ 2.01 & \nodata  &  \nodata & \nodata  &1.29E+02\\
  IC0342 &56.70208&68.09637&0&0.5&31.4&149.72&0.57&94.84&1.34&43.14&1.31&16.24&1.33&3.65E+04\\
  IC2574 &157.19887&68.46756&4.41&0.5&27.2&0.81&0.18&2.26&0.38& $<$ 1.87 & \nodata  &  \nodata & \nodata  &2.43E+02\\
 NGC0337 &14.95871&-7.57797&0&1.87&20&18.57&0.18&6.75&0.31& $<$ 0.88 & \nodata  & $<$ 2.24 & \nodata  &1.69E+03\\
 NGC0628 &24.18846&15.79639&2.37&0.7&20&6.77&0.13&2.19&0.28& $<$ 1.58 & \nodata  & $<$ 4.71 & \nodata  &5.40E+02\\
 NGC0855 &33.51454&27.87733&0&0.94&20&5.82&0.14&2.53&0.26& $<$ 1.17 & \nodata  &  \nodata & \nodata  &4.93E+02\\
 NGC0925 &36.82034&33.57917&0&0.88&20&11.13&0.15&1.31&0.3& $<$ 0.60 & \nodata  &  \nodata & \nodata  &6.50E+02\\
 NGC1097 &41.57938&-30.27489&0&1.38&20&79.91&0.1&36.2&0.26&23.44&0.12&14.42&0.42&1.91E+04\\
 NGC1266 &49.00312&-2.42736&0&2.97&20&6.73&0.2&3.37&0.33&0.78&0.21& $<$ 5.37 & \nodata  &3.91E+03\\
 NGC1291 &49.32746&-41.10806&0&1.01&20&2.62&0.13&0.87&0.29& $<$ 1.07 & \nodata  &  \nodata & \nodata  &5.65E+02\\
 NGC1316 &50.67384&-37.20822&0&2.04&20&4.13&0.23&1.7&0.37& $<$ 0.88 & \nodata  &  \nodata & \nodata  &8.80E+02\\
 NGC1377 &54.16283&-20.90225&0&2.39&20&0.73&0.22& $<$ 1.10 & \nodata  & $<$ 0.65 & \nodata  & $<$ 9.21 & \nodata  &1.33E+03\\
 NGC1404 &54.71633&-35.59439&0&1.96&20& $<$ 0.44 & \nodata  & $<$ 0.92 & \nodata  & $<$ 0.91 & \nodata  &  \nodata & \nodata  &2.37E+00\\
 NGC1482 &58.66233&-20.50267&0&2.19&20&80.74&0.29&42.83&0.5&8.64&0.25& $<$28.48 & \nodata  &1.18E+04\\
 NGC1512 &60.97617&-43.34886&0&1.13&20&10.47&0.24&2.61&0.27&1.23&0.14&  \nodata & \nodata  &1.31E+03\\
 NGC2146 &94.65713&78.35703&0&1.67&20&421.05&1.2&174.6&0.58&50.83&1.46&16.13&1.45&4.49E+04\\
 NGC2798 &139.34497&41.99972&0&2.5&20&32.37&0.27&21.94&0.53&3.24&0.23& $<$ 7.06 & \nodata  &6.50E+03\\
 NGC2841 &140.51096&50.97653&0&1.37&20&3.22&0.11& $<$ 0.79 & \nodata  & $<$ 0.42 & \nodata  &  \nodata & \nodata  &3.47E+02\\
 NGC2915 &141.54803&-76.62634&0&0.5&27.3&5.26&0.24&2.9&0.43& $<$ 1.87 & \nodata  &  \nodata & \nodata  &3.18E+02\\
 NGC2976 &146.78026&67.93239&1.27&0.5&29&31.54&0.34&11.04&0.48& $<$ 0.97 & \nodata  & $<$11.01 & \nodata  &2.69E+03\\
 NGC3049 &148.70648&9.27108&0&1.86&20&9.07&0.13&4.79&0.46&0.98&0.24& $<$ 4.79 & \nodata  &8.85E+02\\
 NGC3077 &150.82947&68.73391&0&0.5&26.9&43.14&0.29&13.67&0.66&2.18&0.4& $<$ 6.23 & \nodata  &4.82E+03\\
 NGC3184 &154.57025&41.42406&0&1.13&20&5.35&0.13&1.44&0.23&1&0.16&  \nodata & \nodata  &7.25E+02\\
 NGC3190 &154.52347&21.83231&0&1.87&20&4.14&0.2&2.13&0.39&0.67&0.15&  \nodata & \nodata  &1.66E+03\\
 NGC3198 &154.97896&45.54961&0&1.37&20&6.48&0.17&1.52&0.36&0.44&0.14&  \nodata & \nodata  &1.07E+03\\
 NGC3265 &157.7782&28.79667&0&1.9&20&7.84&0.18&4.73&0.4& $<$ 0.82 & \nodata  &  \nodata & \nodata  &7.50E+02\\
 NGC3351 &160.9904&11.7038&0&0.9&20&28&0.12&14.03&0.28&6.9&0.24& $<$ 2.82 & \nodata  &6.30E+03\\
 NGC3521 &166.45241&-0.03586&0&1.5&27.6&69.79&0.34&16.04&0.61&5.19&0.63&9.44&2.23&8.45E+03\\
 NGC3621 &169.5688&-32.81406&0&0.64&20&24.2&0.21&5.81&0.36&0.99&0.11&  \nodata & \nodata  &2.28E+03\\
 NGC3627 &170.06888&12.97839&2.39&0.91&20&54.61&0.2&15.78&0.32&6.39&0.31& $<$ 7.81 & \nodata  &6.20E+03\\
 NGC3773 &174.55367&12.11206&0&1.2&20&6.76&0.14&1.96&0.31& $<$ 1.02 & \nodata  &  \nodata & \nodata  &4.88E+02\\
 NGC3938 &178.20604&44.12072&0&1.74&20&9.35&0.2&2.4&0.33&1.33&0.18&  \nodata & \nodata  &1.06E+03\\
 NGC4254 &184.70667&14.4165&0&1.4&20&43.46&0.25&10.45&0.42&5.56&0.21& $<$ 8.88 & \nodata  &5.10E+03\\
 NGC4321 &185.72845&15.82181&0&1.39&20&34.14&0.2&11.2&0.3&6.11&0.22& $<$ 4.53 & \nodata  &5.40E+03\\
 NGC4536 &188.6127&2.18814&0&1.41&20&74.29&0.26&46.55&0.32&4.44&0.33& $<$ 6.74 & \nodata  &9.30E+03\\
 NGC4559 &188.99391&27.95828&0.45&0.68&20&10.79&0.04&2.98&0.08&0.72&0.05&  \nodata & \nodata  &9.60E+02\\
 NGC4569 &189.20746&13.16294&0&0.96&20&14.11&0.16&8.21&0.4& $<$ 0.83 & \nodata  & $<$ 9.86 & \nodata  &2.68E+03\\
 NGC4579 &189.43134&11.81819&0&1.59&20&8.21&0.2&5.01&0.31&1.12&0.13&  \nodata & \nodata  &1.15E+03\\
 NGC4594 &189.99762&-11.62306&0&0.88&20&3.73&0.2& $<$ 1.12 & \nodata  &0.9&0.14&  \nodata & \nodata  &5.75E+02\\
 NGC4625 &190.46967&41.27397&0&0.9&20&7.31&0.15&1.64&0.24& $<$ 1.25 & \nodata  &  \nodata & \nodata  &5.20E+02\\
 NGC4631 &190.53337&32.5415&0&0.74&20&86.7&0.29&25.12&0.33&4.98&0.44& $<$ 5.56 & \nodata  &9.05E+03\\
 NGC4725 &192.61075&25.5008&0&1.15&20&2.02&0.11& $<$ 0.81 & \nodata  & $<$ 0.81 & \nodata  &  \nodata & \nodata  &3.87E+02\\
 NGC4736 &192.73541&41.11958&0.88&0.5&22.1&42.46&0.22&13.83&0.46&2.79&0.28& $<$ 4.22 & \nodata  &5.15E+03\\
 NGC4826 &194.18184&21.68297&0&0.51&20&52.97&0.26&18.98&0.33&11.21&0.43&6.55&0.75&1.42E+04\\
 NGC5055 &198.95554&42.02928&0&0.77&20&32.75&0.14&7.85&0.27&2.99&0.42&7.25&2.29&7.30E+03\\
 NGC5398 &210.3345&-33.06986&1.01&0.74&20&4.4&0.08& $<$ 1.20 & \nodata  & $<$ 0.62 & \nodata  &  \nodata & \nodata  &3.67E+02\\
 NGC5408 &210.82501&-41.37986&0.78&0.5&21.5&5.03&0.23&4.51&0.41& $<$ 1.27 & \nodata  &  \nodata & \nodata  &4.38E+02\\
 NGC5457 &210.80226&54.34894&0&0.65&20&12.22&0.14&3.85&0.2&1.22&0.18&5.27&1.13&1.49E+03\\
 NGC5474 &211.2567&53.66222&0&0.66&20&2.62&0.11& $<$ 0.74 & \nodata  & $<$ 1.51 & \nodata  &  \nodata & \nodata  &2.15E+02\\
 NGC5713 &220.04797&-0.28897&0&2.08&20&52.95&0.17&26.7&0.45&4.88&0.21&5.48&0.91&6.90E+03\\
 NGC5866 &226.62291&55.76322&0&1.48&20&6.51&0.12&1.68&0.36& $<$ 1.23 & \nodata  &3.53&0.75&3.13E+03\\
 NGC6946 &308.71802&60.15392&0&0.66&20&82.76&0.33&40.79&0.32&8.52&0.36&7.43&1.28&1.83E+04\\
 NGC7331 &339.26672&34.41553&0&1.5&21.3&43.65&0.22&10.12&0.47&5.21&0.26& $<$ 2.71 & \nodata  &6.00E+03\\
 NGC7793 &359.45764&-32.59103&0&0.5&26.4&15.37&0.23&4.37&0.51& $<$ 0.88 & \nodata  &  \nodata & \nodata  &1.31E+03\\
\enddata
\label{KFtable}
\end{deluxetable*}
\clearpage
\end{turnpage}
\begin{deluxetable*}{ lcccccccc }[H]
\tablecaption{KINGFISH \cii and CO(2-1) Data}
\tablecomments{Data from the KINGFISH sample.  All errors listed are 1 sigma uncertainties based on the noise in the spectra.  The error for the \cii/CO ratio does not include any absolute calibration uncertainties.} %Brief caption (title)
\tablehead{
 \colhead{Galaxy} & \colhead{Radius} & \colhead{Radius} & \colhead{\cii Flux} & \colhead{\cii $\sigma$} & \colhead{CO 2-1 Flux} & \colhead{CO 2-1 $\sigma$} & \colhead{\cii/CO(2-1)} &\colhead{\cii/CO(2-1)$\sigma$}\\
 \colhead{} & \colhead{(arcsec)} & \colhead{(kpc)} & \multicolumn{2}{c}{(10$^{-16}$ W m$^{-2}$)} & \multicolumn{2}{c}{(10$^{-18}$ W m$^{-2}$)} & \colhead{} & \colhead{} } %Column headings

\startdata

   NGC 0337 &    36.4 &     3.4 &     35.2 &      0.9 &     1.87 &     0.10 &     1885 &  108 \\
   NGC 0628 &   243.3 &     8.5 &     68.2 &      3.6 &    13.03 &     0.24 &      524 &   29 \\
   NGC 0925 &   107.5 &     4.8 &     31.1 &      1.0 &     1.90 &     0.17 &     1633 &  156 \\
   NGC 2146 &    38.0 &     3.2 &    585.2 &      2.4 &    49.70 &     0.14 &     1178 &    6 \\
   NGC 2798 &    36.6 &     4.6 &     36.0 &      0.6 &     8.65 &     0.17 &      416 &   11 \\
   NGC 2841 &   150.9 &    10.3 &     47.7 &      2.5 &     7.25 &     0.86 &      657 &   85 \\
   NGC 2976 &   154.0 &     2.7 &    113.3 &      2.2 &    10.92 &     0.35 &     1037 &   39 \\
   NGC 3049 &    37.6 &     3.5 &     11.2 &      0.4 &     0.80 &     0.09 &     1398 &  161 \\
   NGC 3077 &    38.1 &     0.7 &     46.4 &      0.4 &     2.00 &     0.07 &     2316 &   78 \\
   NGC 3184 &   181.3 &    10.3 &     35.7 &      1.5 &     9.04 &     0.22 &      395 &   19 \\
   NGC 3198 &    89.9 &     6.2 &     19.2 &      1.8 &     3.10 &     0.14 &      620 &   66 \\
   NGC 3351 &   144.2 &     6.5 &     45.5 &      1.9 &    14.14 &     0.39 &      321 &   16 \\
   NGC 3521 &   200.6 &    10.9 &    255.8 &      2.7 &    35.67 &     0.62 &      717 &   15 \\
   NGC 3627 &   136.5 &     6.2 &    227.0 &      7.2 &    42.92 &     0.76 &      529 &   19 \\
   NGC 3938 &   150.5 &    13.1 &     69.1 &      2.7 &     9.18 &     0.21 &      753 &   34 \\
   NGC 4254 &   122.5 &     8.6 &    222.6 &      6.9 &    42.80 &     0.36 &      520 &   17 \\
   NGC 4321 &   166.7 &    11.6 &    109.1 &      3.1 &    34.20 &     0.29 &      319 &    9 \\
   NGC 4536 &   144.5 &    10.2 &    125.9 &      3.0 &    18.29 &     0.52 &      689 &   26 \\
   NGC 4559 &   134.4 &     4.5 &     49.6 &      0.5 &     2.19 &     0.27 &     2265 &  283 \\
   NGC 4569 &   101.5 &     4.9 &     24.0 &      1.2 &    18.74 &     0.28 &      128 &    7 \\
   NGC 4579 &   131.3 &    10.4 &     24.3 &      1.7 &    11.88 &     0.43 &      204 &   16 \\
   NGC 4625 &    37.2 &     1.7 &     12.1 &      0.6 &     0.56 &     0.08 &     2175 &  324 \\
   NGC 4631 &   294.8 &    10.9 &    461.3 &      5.1 &    44.99 &     0.61 &     1025 &   18 \\
   NGC 4725 &    40.4 &     2.3 &      3.7 &      0.5 &     2.25 &     0.15 &      164 &   26 \\
   NGC 4736 &   174.0 &     3.9 &    199.3 &      4.5 &    36.83 &     0.62 &      541 &   15 \\
   NGC 6946 &   284.4 &     9.4 &    525.8 &     21.6 &   128.16 &     0.83 &      410 &   17 \\
   NGC 5055 &   215.3 &     8.3 &    207.1 &      8.3 &    51.30 &     0.68 &      404 &   17 \\
   NGC 5457 &   294.8 &     9.6 &    116.5 &      6.3 &    22.32 &     0.50 &      522 &   31 \\
   NGC 5713 &    65.2 &     6.8 &     99.2 &      3.6 &    18.43 &     0.16 &      538 &   20 \\
   NGC 7331 &    89.1 &     6.3 &    111.2 &      1.7 &    17.43 &     0.42 &      638 &   18

\enddata
\label{KFtable2}
\end{deluxetable*}

\setcounter{figure}{0}
\renewcommand{\thefigure}{A\arabic{figure}}
\begin{figure}
\epsscale{1}
\plotone{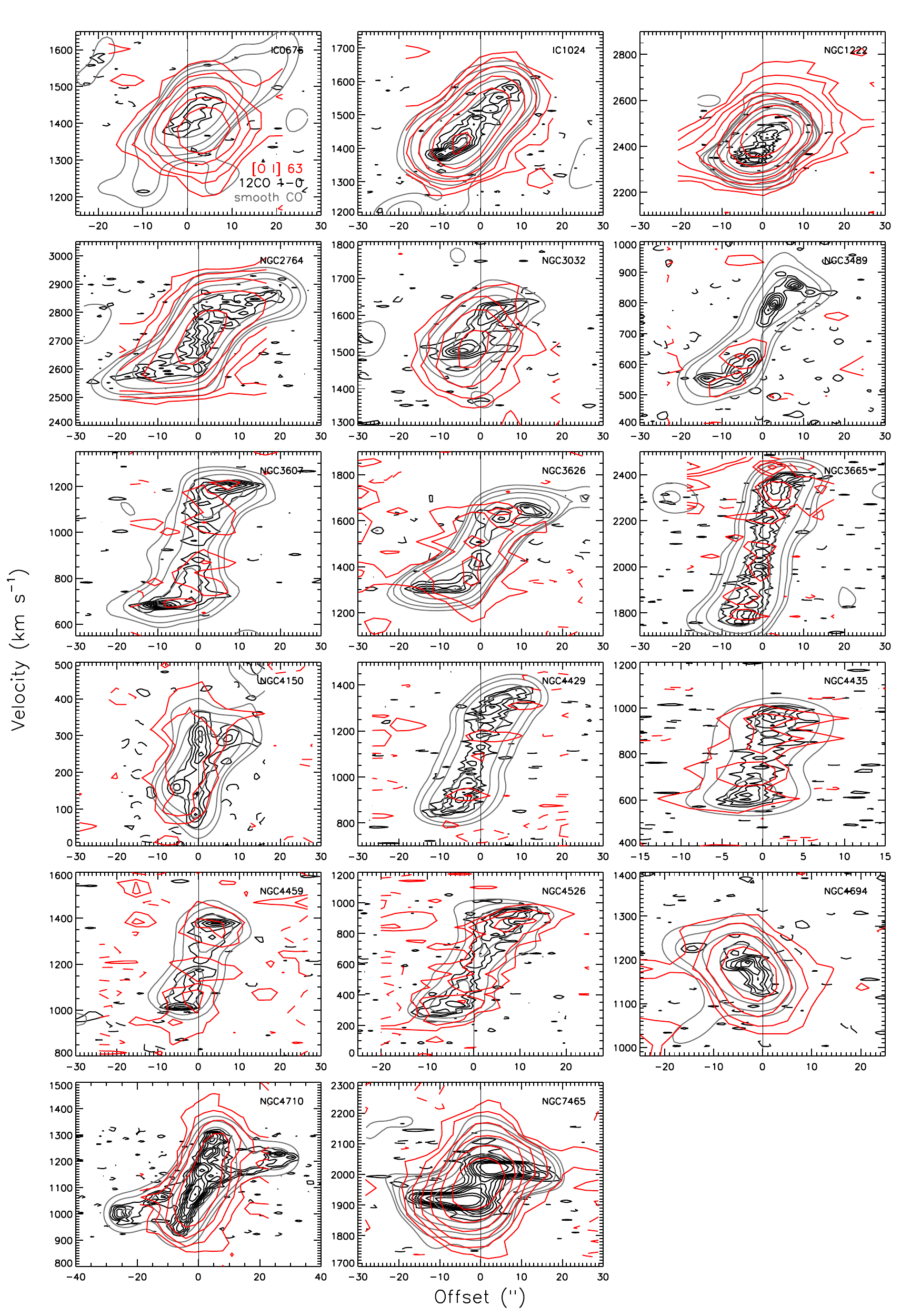} %Plots one file, image width scaled to text width
\caption{\oi position-velocity slices through the optical nucleus, along the kinematic major axis as determined from the CO data (\citet{Paper10}; \citet{Paper16}; \citet{Paper18}).  Velocities on the vertical axis are measured by the optical definition and in the LSR frame, and positions on the horizontal axis are arcseconds (along the major axis) relative to the optical galaxy center at 0.  The Herschel PACS data are shown in red contours, which are drawn at (-2,2,4,6,8,10,12,16,20) times the rms noise level in the center of the cube; the rms noise levels are about 5 mJy/pix in \cii, 2 to 3 mJy/pix in \nii, and 8 to 12 mJy/pix in \oi.  The negative contour is dashed.  For comparison, we also show the CO emission in black contours and, in dark grey contours, the CO emission smoothed to approximately the same spatial and velocity resolution as the PACS data.  Smoothed CO emission is arbitrarily scaled to have the same peak value and same contour levels as the PACS slice.}
\label{pvslice1}
\end{figure}

\begin{figure}
\epsscale{1}
\plotone{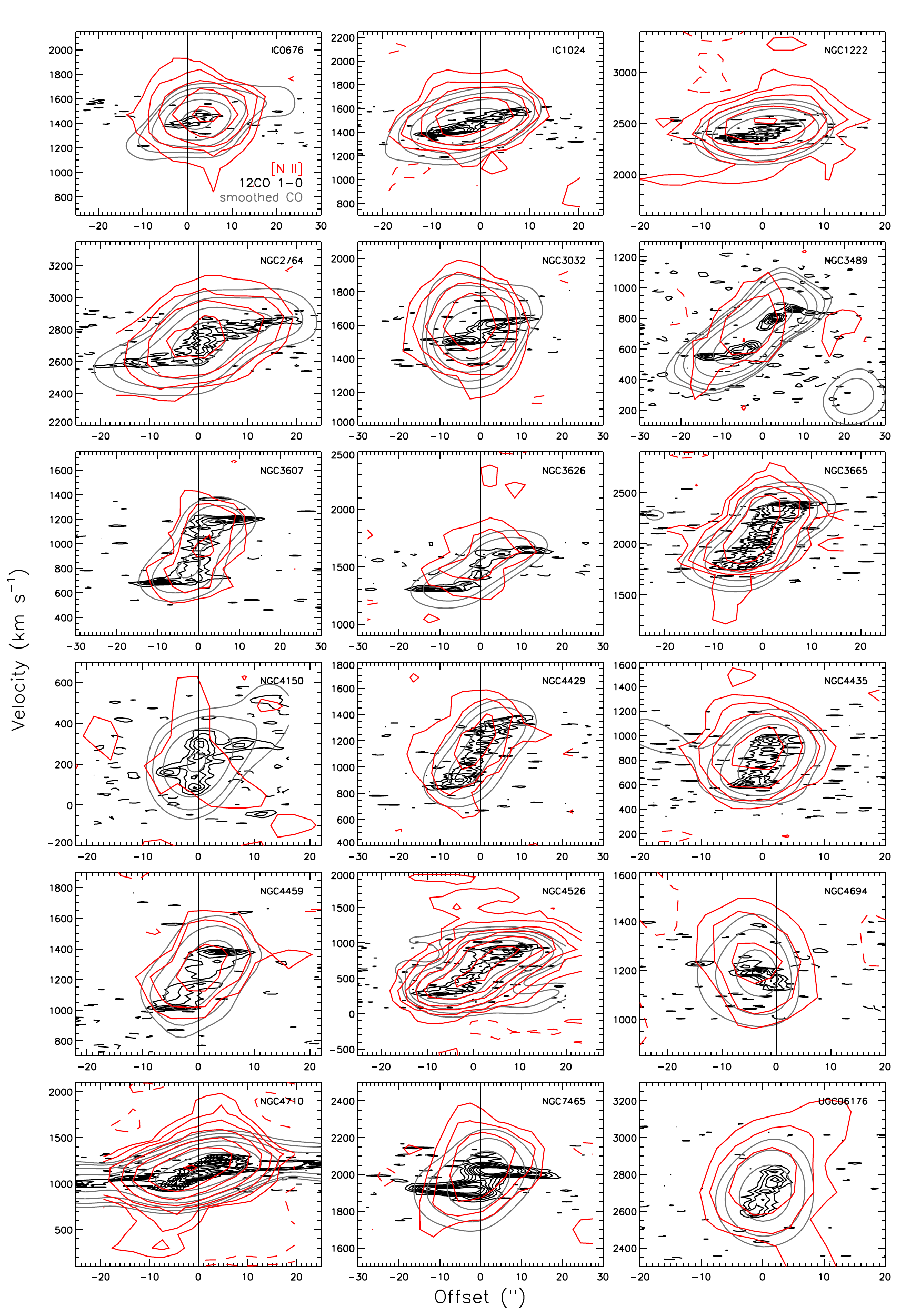} %Plots one file, image width scaled to text width
\caption{\nii position-velocity slices through the optical nucleus, along the kinematic major axis as determined from the CO data (\citet{Paper10}; \citet{Paper16}; \citet{Paper18}).  Velocities on the vertical axis are measured by the optical definition and in the LSR frame, and positions on the horizontal axis are arcseconds (along the major axis) relative to the optical galaxy center at 0.  The Herschel PACS data are shown in red contours, which are drawn at (-2,2,4,6,8,10,12,16,20) times the rms noise level in the center of the cube; the rms noise levels are about 5 mJy/pix in \cii, 2 to 3 mJy/pix in \nii, and 8 to 12 mJy/pix in \oi.  The negative contour is dashed.  For comparison, we also show the CO emission in black contours and, in dark grey contours, the CO emission smoothed to approximately the same spatial and velocity resolution as the PACS data.  Smoothed CO emission is arbitrarily scaled to have the same peak value and same contour levels as the PACS slice.}
\label{pvslice2}
\end{figure}

\begin{figure}
\epsscale{1}
\plotone{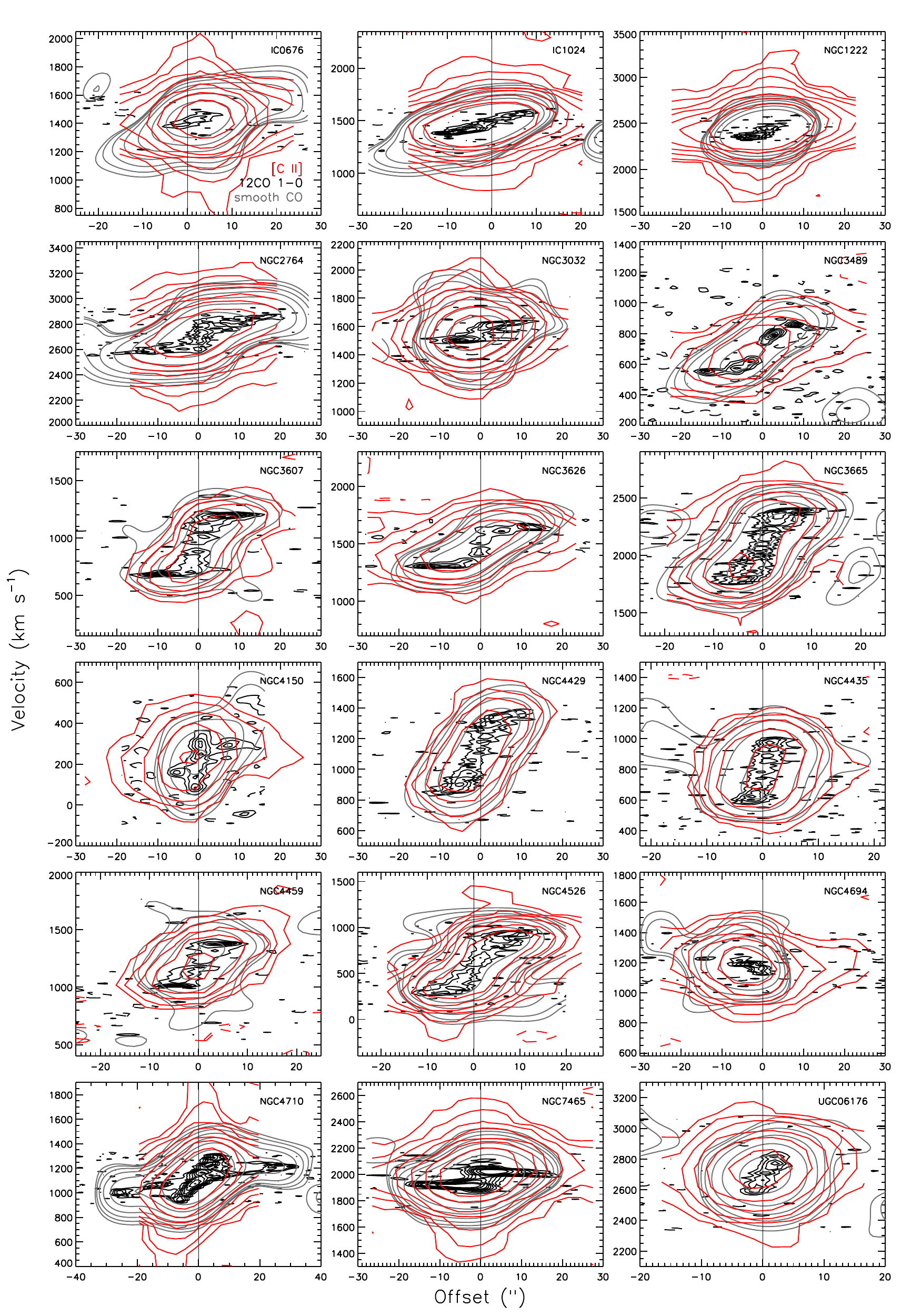} %Plots one file, image width scaled to text width
\caption{\cii position-velocity slices through the optical nucleus, along the kinematic major axis as determined from the CO data (\citet{Paper10}; \citet{Paper16}; \citet{Paper18}).  Velocities on the vertical axis are measured by the optical definition and in the LSR frame, and positions on the horizontal axis are arcseconds (along the major axis) relative to the optical galaxy center at 0.  The Herschel PACS data are shown in red contours, which are drawn at (-2,2,4,6,8,10,12,16,20) times the rms noise level in the center of the cube; the rms noise levels are about 5 mJy/pix in \cii, 2 to 3 mJy/pix in \nii, and 8 to 12 mJy/pix in \oi.  The negative contour is dashed.  For comparison, we also show the CO emission in black contours and, in dark grey contours, the CO emission smoothed to approximately the same spatial and velocity resolution as the PACS data.  Smoothed CO emission is arbitrarily scaled to have the same peak value and same contour levels as the PACS slice.}
\label{pvslice3}
\end{figure}

\begin{figure}
\epsscale{1}
\plotone{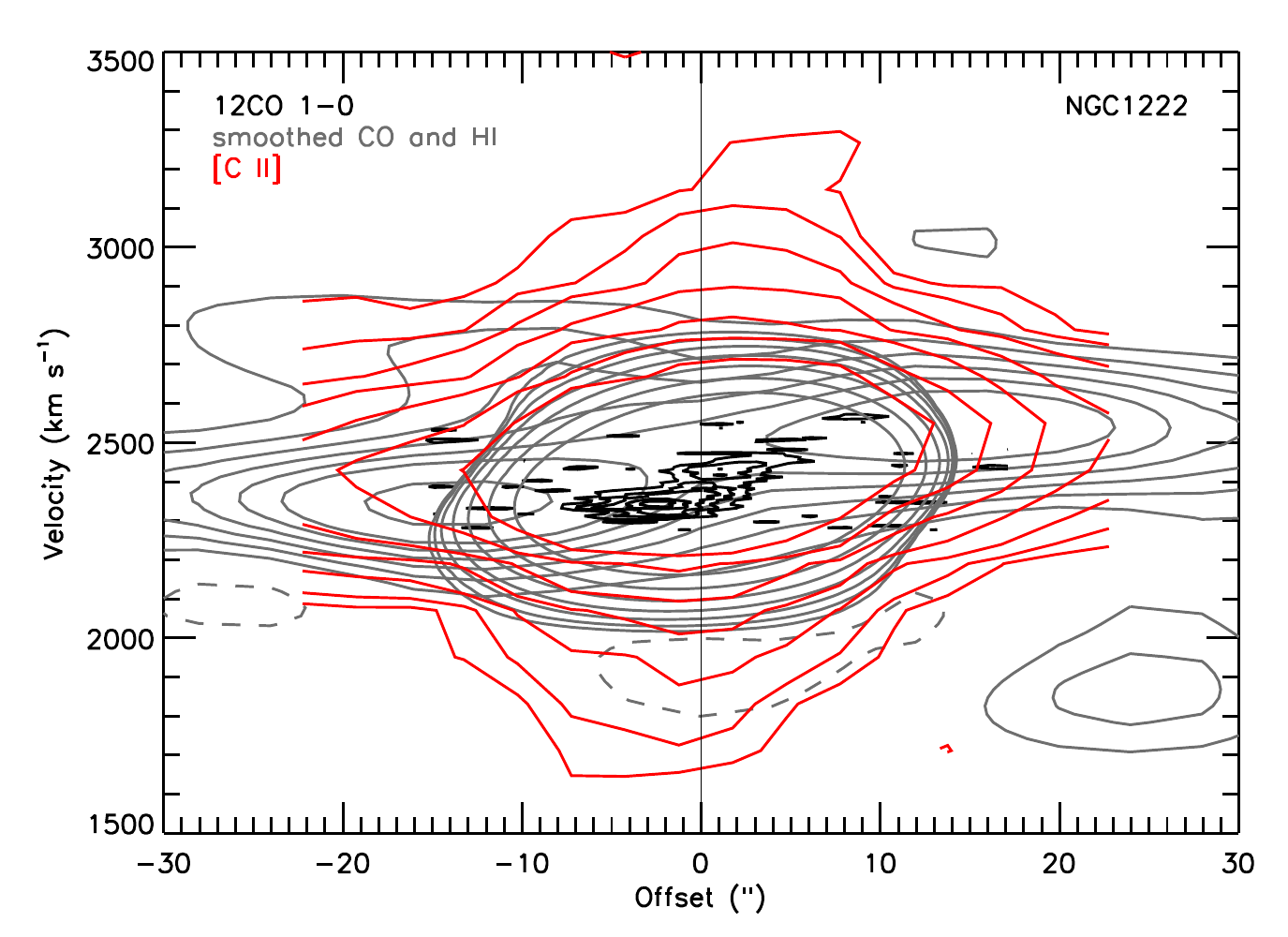}
\caption{In the case of NGC 1222, the \cii emission is clearly more spatially extended than the CO.  This Figure is similar to Figures \ref{pvslice1}, \ref{pvslice2}, and \ref{pvslice3},  but has \hi in dark grey contours. Here the \cii emission probably follows the sum of atomic and molecular gas.  There are other hints that \cii is a bit more extended than CO in NGC 4694, 7465, and IC 1024 as well.}
\label{pv1222}
\end{figure}

\begin{figure}
\hbox{
\includegraphics[scale=0.6,trim=5mm 5mm 0mm 0mm]{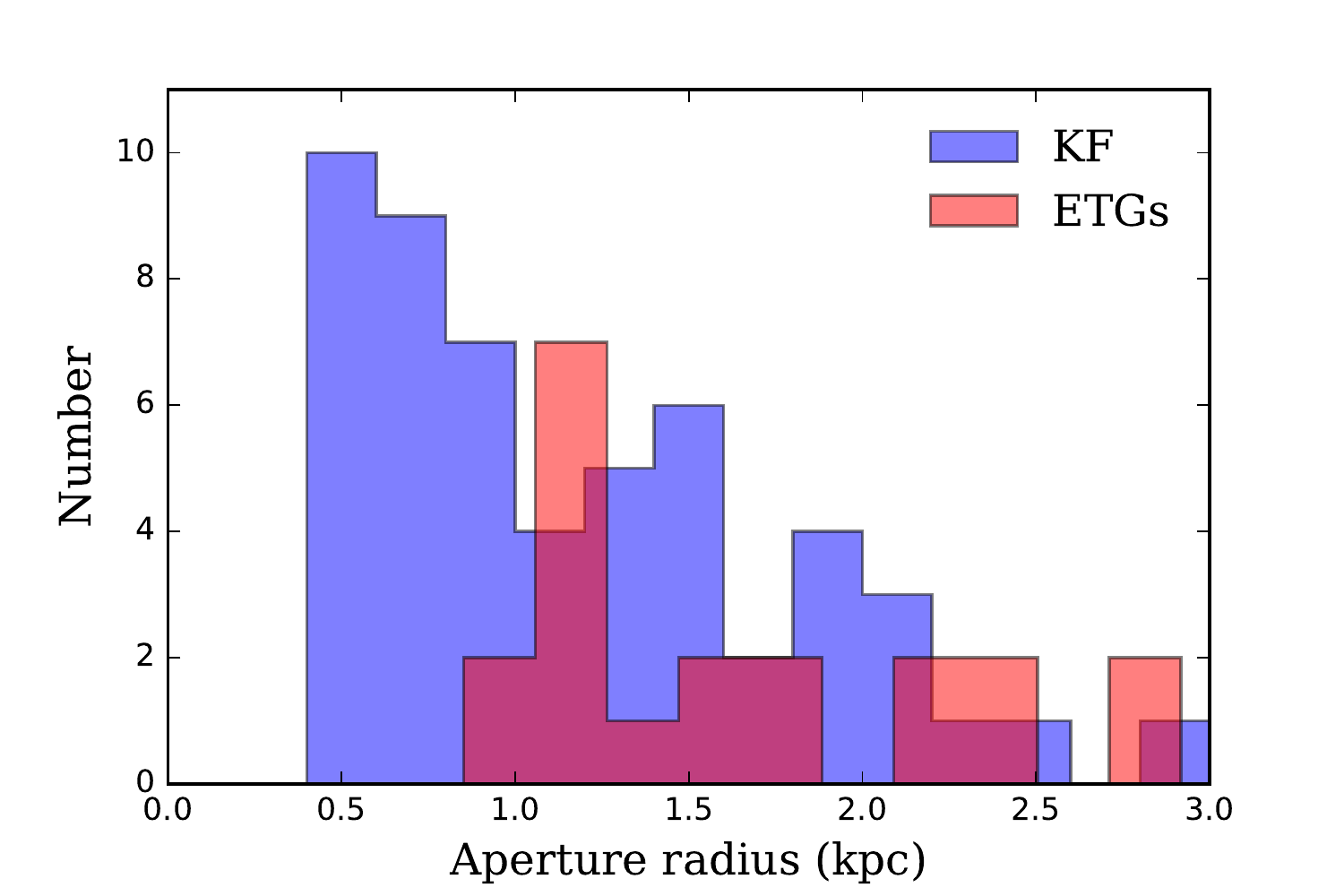}
\includegraphics[scale=0.6,trim=5mm 5mm 0mm 0mm]{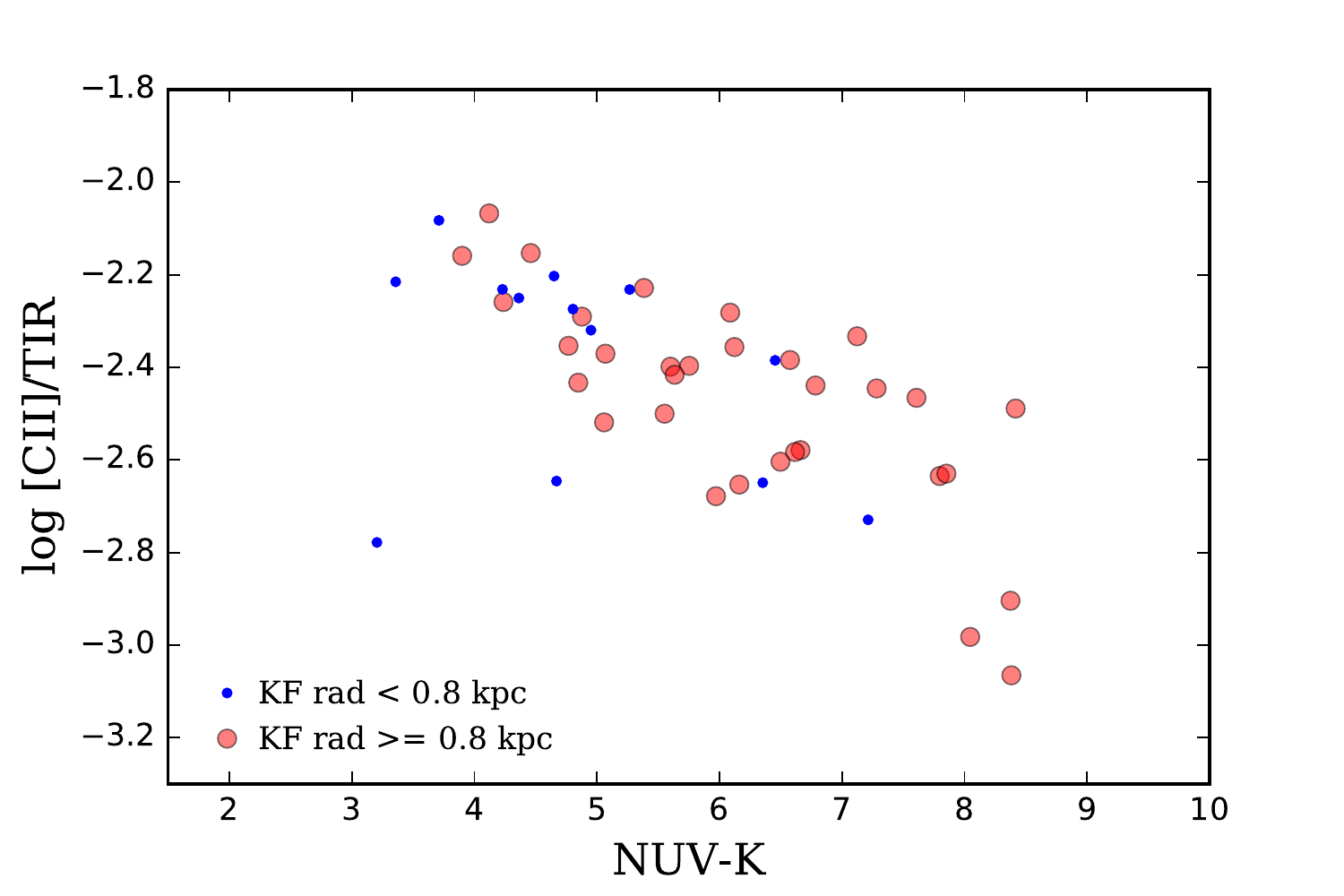}
}
\caption{Left: Size distribution of the circular apertures used for KINGFISH galaxies and our sample of early-type galaxies.  Right: \cii/TIR values versus global NUV-K colors in the KINGFISH sample, with symbols indicating each aperture's linear size.  The larger red points correspond to apertures overlapping the early-type galaxy size distribution.  The smaller blue points tend to be dwarf galaxies closer than $\approx$ 8 Mpc, with limited PACS fields of view.
\label{fig:apertures}}
\end{figure}

\end{appendix}
%\section{}
%
%%%%============== End Appendix =========================

%%%---------------- Bibliography --------------------------
\FloatBarrier
\newpage
\bibliographystyle{aj}

%%%============== End Bibliography =========================

\end{document}